\documentclass[twoside,leqno,twocolumn]{article}

\usepackage[english]{babel}

\usepackage{amsmath}
\usepackage{amssymb}
\usepackage{amsthm}
\usepackage{graphicx}
\usepackage[colorlinks=true, allcolors=blue]{hyperref}

\usepackage[linesnumbered]{algorithm2e}
\usepackage{algpseudocode}

\usepackage{fancyhdr}

\usepackage{ifthen}

\usepackage[final]{changes}

\definechangesauthor[name=Review01, color=olive]{R01}
\definechangesauthor[name=Review02, color=blue]{R02}
\definechangesauthor[name=Review03, color=red]{R03}
\definechangesauthor[name=Us, color=brown]{us}

\begin{document}

\title{\Large 
 SYCL compute kernels for ExaHyPE\relatedversion
}

\author{
 Chung Ming Loi\thanks{
  Department of Computer Science, Durham University, United Kingdom
%   \email{chung.m.loi@durham.ac.uk}
  \href{mailto:chung.m.loi@durham.ac.uk}{chung.m.loi@durham.ac.uk}
 }
 \and
 Heinrich Bockhorst\thanks{
  Intel Deutschland GmbH, Germany
%   \email{tobias.weinzierl@durham.ac.uk}
  \href{mailto:heinrich.bockhorst@intel.com}{heinrich.bockhorst@intel.com}
 }
 \and 
 Tobias Weinzierl\thanks{
  Department of Computer Science, Institute for Data Science, Durham University,
  United Kingdom
%   \email{tobias.weinzierl@durham.ac.uk}
  \href{mailto:tobias.weinzierl@durham.ac.uk}{tobias.weinzierl@durham.ac.uk}
 }
}
\newcommand\relatedversion{}
\renewcommand\relatedversion{\thanks{The full version of the paper can be accessed at \protect\url{https://arxiv.org/abs/2306.16731}}} % Replace URL with link to full paper or comment out this line

\date{}

\maketitle

% Copyright Statement
% When submitting your final paper to a SIAM proceedings, it is requested that you include
% the appropriate copyright in the footer of the paper.  The copyright added should be
% consistent with the copyright selected on the copyright form submitted with the paper.
% Please note that "20XX" should be changed to the year of the meeting.

% Default Copyright Statement
\fancyfoot[R]{\scriptsize{Copyright \textcopyright\ 20XX by SIAM\\
Unauthorized reproduction of this article is prohibited}}

\newtheorem{definition}  {Definition}
\newtheorem{observation} {Observation}
\newtheorem{workaround}  {Idiom}

\DontPrintSemicolon
\newenvironment{bash}[1][htb]{
  \renewcommand{\algorithmcfname}{Bash commands}
  \begin{algorithm}[#1]%
}
{\end{algorithm}}

\begin{abstract}
 \small\baselineskip=9pt
 We discuss three SYCL realisations of a simple Finite Volume scheme over
multiple Cartesian patches. 
The realisation flavours differ in the way how they map the compute steps onto
loops and tasks:
We compare an implementation \replaced[id=R01]{that}{which} is exclusively
using a \replaced[id=us]{sequence}{cascade} of for-loops to a version
\replaced[id=R01]{that}{which} uses nested parallelism, and finally benchmark these against a version \replaced[id=R01]{modelling}{which models}  the calculations as task
graph.
Our work proposes realisation idioms to realise these flavours within SYCL.
The \deleted[id=us]{idioms translate to some degree to other GPGPU programming
techniques, too.
Our preliminary} results suggest that a mixture of classic task and data
parallelism performs \replaced[id=us]{if}{best, as long as} we map this hybrid
onto a solely data-parallel SYCL implementation\added[id=us]{, taking into
account SYCL specifics and the problem size}.

\end{abstract}

\section{Introduction}
\label{section:introduction}

%\marginpar{Tom,Michael,Mario,Rod,Intel,Dominic,alle Konsortien,Michael Wong}

%
% Big context
%
Exascale and pre-exascale pioneers obtain the majority of their compute power
from GPGPUs.
Their nodes are heterogeneous, with a general-purpose CPU, the host, being
supplemented by streaming multiprocessor cards.
Delivering code for such architectures is challenging, once we want to
harness both the capabilities of the CPU cores and the GPUs.
We need code \replaced[id=R01]{that}{which} runs on either compute platform.
In our work, we \replaced[id=us]{study}{focus on} a block-structured Finite
Volume code for wave equations\deleted[id=R01]{, which has to be ported to GPU-accelerated clusters}.

%
% Loop paradigm is key
%
Performance analysis shows that the majority of its runtime is spent in
one compute routine which we call a computational kernel.
It takes a set of small Cartesian meshes of Finite Volumes (blocks\added[id=us]{
or patches}) and advances them in time
\cite{Li:2022:TaskFusion,Wille:2023:GPUOffloading}.
This kernel \replaced[id=us]{spans a 
 small, static execution graph over nested loops.
}{
 consists of nested loops and spans a 
 small, static execution graph.
} 
Making the kernel perform on GPGPUs and CPUs is key
to exploit heterogeneous systems.
We assume that this is typical for many scientific codes.

%
% Challenge
% 
Programmers have various technologies at hand to realise such compute kernels.
OpenMP \cite{Klemm:2021:OpenMP}, SYCL \cite{Reinders:2021:DPC++} and Kokkos
\cite{Trott:2022:Kokkos} are popular examples.
Our work here focuses on SYCL. 
However, we address fundamental questions \replaced[id=R01]{that}{which} also apply---partially and/or
to some degree---to our OpenMP \cite{Wille:2023:GPUOffloading} and C++ GPGPU
ports of the compute kernel as well as other numerical
schemes within the underlying software \cite{Reinarz:2020:ExaHyPE}:
What patterns and idioms exist to write kernels in languages \replaced[id=R01]{that}{which} support
both data- and task-parallelism and focus on GPUs with their SIMD/SIMT hardware
parallelism?
Further to that, is it possible to make early-day statements on the
expected efficiency impact of using one or combining the two parallelism
paradigms, i.e.~tasks vs data parallelism?

%
% What we do
%
We start from the definition of a microkernel \replaced[id=R01]{that}{which} is embedded into
the kernel's compute blueprint. 
Microkernels represent invocations of domain-specific code fragments.
In the Finite Volume context this includes the flux and eigenvalue calculations
for example.
Through microkernels, we keep the domain-specific code separate from the
way how the kernel is realised. 
We next identify three different approaches how to map the arising compute
graph, which is a graph over the invocation of user functions aka microkernels,
and its required (temporary) data structures onto SYCL kernels.
For these approaches, we discuss realisation idioms.

%
% Contextualisation. Why is is so cool and why is it novel aka worth reading
%
Our research stands in the tradition of work \replaced[id=R01]{that}{which} distinguishes the role of the
performance engineer strictly from the role of domain scientists, numerics
experts, research software engineers, and further
specialists \replaced[id=us]{participating in}{contributing towards} the
computational sciences workflow \cite{Gallard:2022:Roles}.
We focus on \replaced[id=us]{the performance aspect}{performance} in a GPGPU
context.
As our approach never alters domain code, traditional performance tuning
opportunities targeting the calculations or data layout are largely off the
table.
Instead, we have to focus solely on the orchestration of concurrency.
Such work closes a gap between the formulation of an algorithm and its
\replaced[id=us]{realisation in SYCL}{
optimisation once it is translated into a cascade of loops or the
application of matrix applications for example}.
To the best of our knowledge, such work is largely absent in literature.

%
% SUmmary of key insights and shortcomings
%
Our data suggest that flexible, high-level task parallelism 
\replaced[id=us]{underperforms relative}{continues to
underperform compared}  to plain loop parallelism.
However, it is not clear if this is due to the newness of the
underlying software stack or indeed an intrinsic consequence of the underlying
SIMT hardware.
\replaced[id=R3]{
 Our measurements also suggest that SYCL's Unified Shared Memory (USM) 
 is slow and that codes benefit from explicit copying or
 host-managed GPU memory as introduced in \cite{Wille:2023:GPUOffloading}.
 We
}{
We also emphasise that we do not include data transfer in our considerations.
Though efficient data migration strategies \cite{Wille:2023:GPUOffloading} are
key to exploit GPGPUs efficiently, we
} 
note that future hardware generations
might actually ``solve'' this by a tight GPU-CPU integration.
This implies that efficient compute kernel realisations gain even more
importance.
Finally, we focus on two GPU generations only and hence omit any
discussion to which degree our realisation flavours are performance portable
\cite{Deakin:2022:HeterogeneousProgramming}.

%
% Structure
%
The remainder is organised as follows:
We first introduce our underlying simulation code and formalise its Finite
Volume numerics via a task graph over microkernels
(Sec.~\ref{section:finite-volumes}).
The main body of work in Section \ref{section:realisations} discusses three
approaches how to map the computational scheme onto a kernel implementation.
We continue with a discussion of the implementation of these schemes within SYCL
(Sec.~\ref{section:sycl}), before we present some results in
Sec.~\ref{section:results}.
After a reflection on the lessons learned\deleted[id=R01]{
(Sec.~\ref{section:evaluation})}, a brief outlook \replaced[id=R01]{closes}{and
conclusion close} the discussion.
Our work is complemented by instructions how to reproduce all results
(Appendices~\ref{appendix:download-and-compilation} and
\ref{appendix:execution-and-postprocessing}), as well as a section presenting further data.

\section{ExaHyPE's Finite Volumes}
\label{section:finite-volumes}

Our work employs ExaHyPE 2, an engine to solve hyperbolic equation systems

\vspace{-0.2cm}
\begin{equation}
 \frac{\partial}{\partial t} Q
 +
 \boldsymbol{\nabla}\cdot\textbf{F}(Q)
 +
 \sum_i \boldsymbol{ \mathcal{B} }_i\,\frac{\partial Q}{\partial x_i}
 = 
   \textbf{S}(Q)
 +
    \sum \delta,
    \label{equation:finite-volumes:PDE}
\end{equation}
\vspace{-0.4cm}

\noindent
given in first order formulation.
ExaHyPE 2 is a rewrite of the first-generation ExaHyPE
\cite{Reinarz:2020:ExaHyPE} \replaced[id=R01]{that}{which} in turn employs principles advocated for in the underlying
\replaced[id=us]{adaptive meshing}{AMR} framework Peano
\cite{Weinzierl:2019:Peano}:
Users specify the number of unknowns held by $Q \in \mathbb{R}^N$ and provide
implementations of the the numerical terms $\textbf{F}_n(Q), \boldsymbol{ \mathcal{B} }_{i,n}, S,
\delta:
\mathbb{R}^{N} \mapsto \mathbb{R}^{N}$\replaced[id=R01]{. These}{which}
represent (conservative) flux, non-conservative fluxes, volumetric and point sources.
The software then
automatically assembles a solver for the equation system from
(\ref{equation:finite-volumes:PDE}).
As we stick strictly to Cartesian formulations, the software
expects directional fluxes along \replaced[id=us]{the axes}{an axis $n$}.

\subsection{Block-structured adaptive mesh refinement with tasks}

Our code provides various explicit time stepping schemes to solve
these PDEs over adaptive Cartesian meshes \deleted[id=R01]{which are}
spanned by spacetrees \cite{Weinzierl:2019:Peano}:
The computational domain is covered by a square or cube \replaced[id=R01]{that}{which} is recursively
subdivided into three equidistant parts along each coordinate axis. 
We end up with a spacetree that spans an adaptive Cartesian grid.
Into each octant of this spacetree grid, we embed a Cartesian $p \times p$
($d=2$) or $p \times p \times p$ ($d=3$) mesh.
This is the actual compute data structure.
It equals, globally, a
block-structured adaptive Cartesian mesh \cite{Dubey:16:SAMR}.

ExaHyPE employs a cascade of
parallelisation techniques:
It splits the spacetree mesh along the Peano space-filling curve into chunks and
distributes these chunks among the ranks.
Each rank employs the same scheme again to keep the cores busy.
This is a classic non-overlapping domain decomposition over the mesh of octants
hosting the Cartesian patches.
In a third step, the code identifies those patches per subdomains \replaced[id=R01]{that}{which} are
non-critical, i.e.~are not placed along the critical path of the execution
graph, and also can be executed non-deterministically.
It deploys their updates onto tasks 
\cite{Charrier:2020:EnclaveTasking,Li:2022:TaskFusion}.

Among these so-called enclave tasks,
ExaHyPE identifies tasks on-the-fly \replaced[id=R01]{that}{which} are free of global side
effects \cite{Li:2022:TaskFusion}, i.e.~do not have to be executed within the node's shared
memory space, bundles $T$ of these tasks into one large meta task assembly, and
then process them via one large compute kernel in one
rush.
Such assemblies can be deployed to GPGPUs \cite{Wille:2023:GPUOffloading}.

\subsection{Kernels and microkernels}

In the present paper, we employ Finite Volumes with a
generic Rusanov solver.
This is the simplest, most generic solver offered through ExaHyPE.
Each volume within the $p^d$ patch carries the $N$ quantities of interest from
(\ref{equation:finite-volumes:PDE}).
For a given solution $Q$ over a volume $c$ and time step size $\Delta t$, we
determine a new \deleted[id=us]{solution}

\vspace{-0.2cm}
\[
 Q |_c \gets Q |_c \pm \Delta t \cdot h \sum _{\partial c}F|_{n} (Q),
\]
\vspace{-0.4cm}

\noindent
where the
flux over each face $\partial c$ with a normal $n \in \{x,y,z\}$ is
\replaced[id=us]{computed}{determined} as

\vspace{-0.4cm}
\begin{eqnarray}
  F|_{n} (Q) & \approx & \frac{1}{2} \left( F_n^{+} (Q) + F_n^{-}(Q) \right)
   - \frac{C}{h} \cdot \left( Q^{+} - Q^{-} \right) 
  \nonumber
  \\
  & & \text{max} (\lambda _{\text{max},n}(Q^{+}), \lambda
  _{\text{max},n}(Q^{-})).
  \label{equation:finite-volumes:Rusanov}
\end{eqnarray}
\vspace{-0.4cm}

\noindent
There is no unique solution $F|_{n} (Q)$ on the face, since $Q$ 
jumps from one volume into the other.
Therefore, we approximate the flux through
the $^+ / ^-$ values in the cells left and right of the
face, where all $N$ quantities are uniquely defined.
We omit the treatment of the asymmetric non-conservative term 
$\boldsymbol{ \mathcal{B} }_i $ in
(\ref{equation:finite-volumes:Rusanov})\added[id=us]{ in the discussion here},
but \added[id=us]{emphasise that we} require the user to provide
implementations $\lambda _{\text{max},n}(Q):
\mathbb{R}^N \mapsto \mathbb{R}$\replaced[id=R01]{. They}{ which} yield the maximum eigenvalue (wave
speed) over the flux for a given quantity along the axis $n$.

To make the scheme work, each patch is supplemented with a halo layer of size
one. 
ExaHyPE \cite{Reinarz:2020:ExaHyPE} and its underlying AMR framework Peano
\cite{Weinzierl:2019:Peano} manage the adaptive Cartesian mesh, distribute it
among the cores, identify tasks, equip the patches associated with these tasks
with a halo layer, and then invoke an update kernel for the patch following
(\ref{equation:finite-volumes:Rusanov}).
Once updated, the kernel eventually might compute the new maximum eigenvalue
over the new solution.
This reduced eigenvalue feeds into the CFL condition and determines the next
time step size.
As we support local time stepping, the maximum eigenvalue has to
be determined per patch.

\subsection{DAG}

Our GPU update kernel accepts $N \cdot (p+2)^d \cdot T$ quantities (double
values) associated with the $T$ patches (including their halo) and yields the $N
\cdot p^d \cdot T$ quantities at the next time step.

\begin{definition} 
 A \emph{microkernel} is a function \replaced[id=R01]{accepting}{which accepts}
 a volume index $ [0,\ldots,T-1] \times [-1,0,\ldots,p]^d$, \linebreak pointers
 to some input and output arrays as well as some meta information such as the time step size. Each microkernel wraps one
 compute step such as a flux evaluation or an update of a cell according to an
 additive term from (\ref{equation:finite-volumes:Rusanov})\deleted[id=us]{ and
 updates the image}.
\end{definition}

\noindent
As a microkernel is passed an index, we can parameterise the microkernels
with a function $enum: [0,\ldots,T-1] \times [-1,0,\ldots,p]^d \mapsto
\mathbb{N}^+_0$.
It encodes how we order the input and output data\added[id=us]{ in memory}.
Therefore, each microkernel call knows exactly what elements from the data
arrays to read and write.
%  one scalar (for $\lambda _{\text{max},n}$) or $N$-valued
% output tied to one volume.

\begin{algorithm}[htb]
 {\footnotesize
 \For{patch $\in [0,\ldots,T-1]$}{
  allocate $tmp_{\text{F}_x},tmp_{\text{F}_y} \in \mathbb{R}^{NT \cdot
  (p+2)^d}$\; 
  allocate $tmp_{\lambda_x},tmp_{\lambda_y} \in \mathbb{R}^{T \cdot
  (p+2)^d}$\; 
  \For{$c \in [0,p-1] \times [0,p-1]$}{
   $Q^{(new)}(patch,c) \gets Q(patch,c)$;
   \Comment microkernel
  }
  \For{$c \in [-1,p] \times [0,p-1]$}{
   $tmp_{\text{F}_x} \gets F_x(Q)(patch,c)$;
   \Comment microkernel
  }
  \For{$c \in [0,p-1] \times [-1,p]$}{
   $tmp_{\text{F}_y} \gets F_y(Q)(patch,c)$;
   \Comment microkernel
  }
  \For{$c \in [0,p-1]^d$}{
   $Q^{(new)}(patch,c) += \Delta t \cdot h \cdot tmp_{\text{F}_x}
   (patch,\text{left of}\ c)$\; 
   $Q^{(new)}(patch,c) -= \Delta t \cdot h \cdot tmp_{\text{F}_x}
   (patch,\text{right of}\ c)$\; 
   \ldots \Comment microkernels
  }
  \For{$c \in [-1,p] \times [0,p-1]$}{
   $tmp_{\lambda_x} \gets \lambda _{\text{max},x}(Q)(patch,c)$;
   \Comment microkernel
  }
  \ldots \;
 }
 }
 \vspace{0.2cm}
 \caption{
   Schematic sketch of a $2d$ compute kernel over a set of patches.
   \label{algorithm:kernel}
 \vspace{-0.4cm}
  }
\end{algorithm}

The overall numerical scheme over $T$ patches that is realised by one kernel
invocation (Alg.~\ref{algorithm:kernel}) \replaced[id=us]{can be written
down as}{equals} a series of (nested) loops over microkernels. 
Some of them are trivial copies, others invoke user functions, while again
others combine the outcomes of the latter and add them to the output.
The $\gets $ operator highlights that all memory address calculations are hidden within the
kernels as we instantiate them with an enumerator $enum(patch,c)$.

\section{Kernel realisations}
\label{section:realisations}

\begin{figure}[htb]
 \begin{center}
  \includegraphics[width=0.4\textwidth]{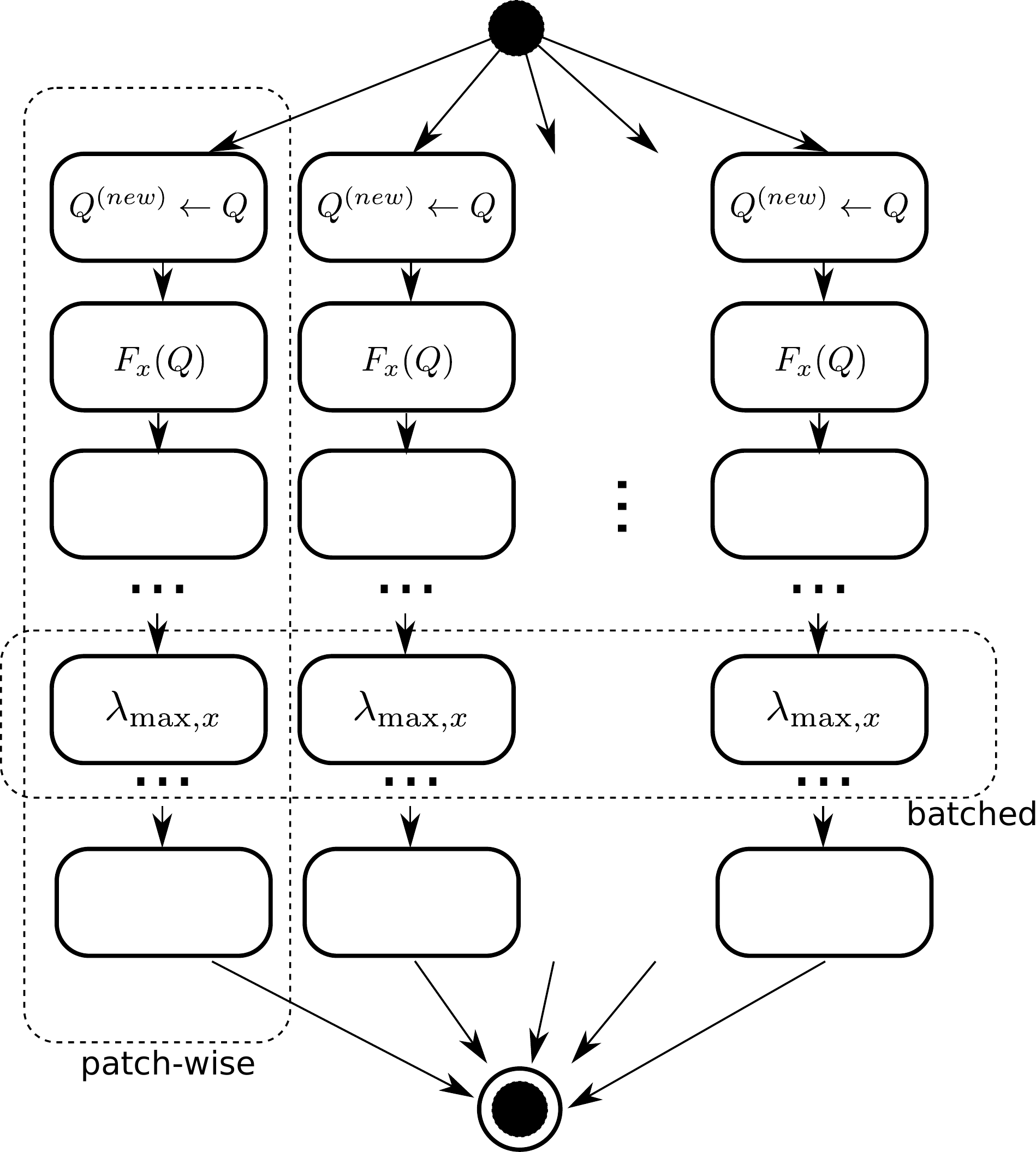}
 \end{center}
 \vspace{-0.6cm}
 \caption{
  Sketch of the compute graph sketch for a kernel over $T$ patches. Each node in
  the graph represents a $d$-dimensional loop over all volumes of the patch
  subject to halo volumes where appropriate.
  \label{figure:realisation:task-graph}
  \vspace{-0.4cm}
 }
\end{figure}

%
% Should be come: In this mansucript, we
%
% Loop-level parallelism is an intrinsic fit to multiprocessors. 
There is a multitude of ways to translate Alg.~\ref{algorithm:kernel} into
SYCL. 
We start from a \replaced[id=us]{rewrite of the algorithm into a directed,
acylic graph (DAG)}{ DAG} over sets of
\replaced[id=us]{calculations}{microkernels} per patch
(Fig.~\ref{figure:realisation:task-graph}).
Each node within this DAG equals a $d$-dimensional loop over 
microkernel calls.
It can be vectorised, i.e.~facilitates coalesced memory access.

% We distinguish three different paradigms to map the remaining DAG with these
% coalesced nodes onto compute kernels:

\begin{definition}
 A \emph{patch-wise} realisation employs a (parallel) outer loop
 \replaced[id=R01]{running}{which runs} over the patches. Within each
 loop ``iteration'', one patch is handled.
\end{definition}

\noindent
Patch-wise kernels group the calculations vertically (Fig.~\ref{figure:realisation:task-graph}).
\replaced[id=us]{An outer loop}{A sequential kernel} handles the set of patches
patch by patch.
While this outer loop yields parallelism over the patches---the SYCL code is
very close to Alg.~\ref{algorithm:kernel} where the outer loop over $patch$ is
running in parallel---there are additional (nested) parallel loops over $c$.
Their ends synchronise the logical steps of
Alg.~\ref{algorithm:kernel} on a per-patch base:
After we have done all the flux calculations along the x-direction for one
patch, we continue with all the calculations along y.
This synchronisation is totally independent of the flux calculations for
any other patch.
Only the very end implements a global synchronisation \replaced[id=R01]{that}{which} coincides with the
end of the loop over the patches.

Globally, the execution graph fans out initially with one branch per patch, 
and it fans in once in the end.
Within each fan branch, the steps run one after\added[id=us]{ the other}.
However, each step fans in and out again.
The concurrency level within the kernel hence always oscillates between $T$
and $\mathcal{O}(T(p+2)^d)$ with only one global synchronisation point in the
end.

\begin{definition}
 A \emph{batched} realisation runs over the logical algorithm steps one-by-one. 
 Within each step, it processes all volumes from all patches in parallel.
\end{definition}

\noindent
This scheme clusters calculations
horizontally (Fig.~\ref{figure:realisation:task-graph}).
It realises the kernel as a \deleted[id=us]{sequential} sequence of algorithmic
steps.
Within each step, we update all volumes from all patches concurrently.
After each step, we synchronise over all patches.
The scheme is labelled as \emph{batched} \cite{Li:2022:TaskFusion}, inspired by
batched linear algebra \cite{Abdelfattah:2020:BatchedBLAS}, while we use the
same (non-linear) operator (matrix) for each patch input $Q$.

Globally, we get a trivial DAG enlisting the calculation steps like pearls in a
row.
Each step within the DAG fans in and and out, i.e.~has an internal
concurrency level in the order $\mathcal{O}(T \cdot (p+2)^d)$.
As we bring the algorithmic steps into an order, i.e.~remove concurrency
from the global DAG by partial serialisation, there are multiple
synchronisation points.

\begin{definition}
 The \emph{task-graph} realisation employs one task graph where each node represents all operations (microkernel calls) of one
 type for one patch.
\end{definition}

\noindent
The task-graph approach is a \replaced[id=us]{plain}{direct} realisation of the
\added[id=us]{logical} task graph
\replaced[id=us]{(Fig.~\ref{figure:realisation:task-graph})}{from
Fig.~\ref{figure:realisation:task-graph}}.
Each individual node within this task graph fans in and out and has an internal
concurrency of $\mathcal{O}((p+2)^d)$, and we may assume that there are always at
least $T$ nodes \deleted[id=us]{within the DAG} ready.

\begin{figure}[htb]
 \begin{center}
  \includegraphics[width=0.4\textwidth]{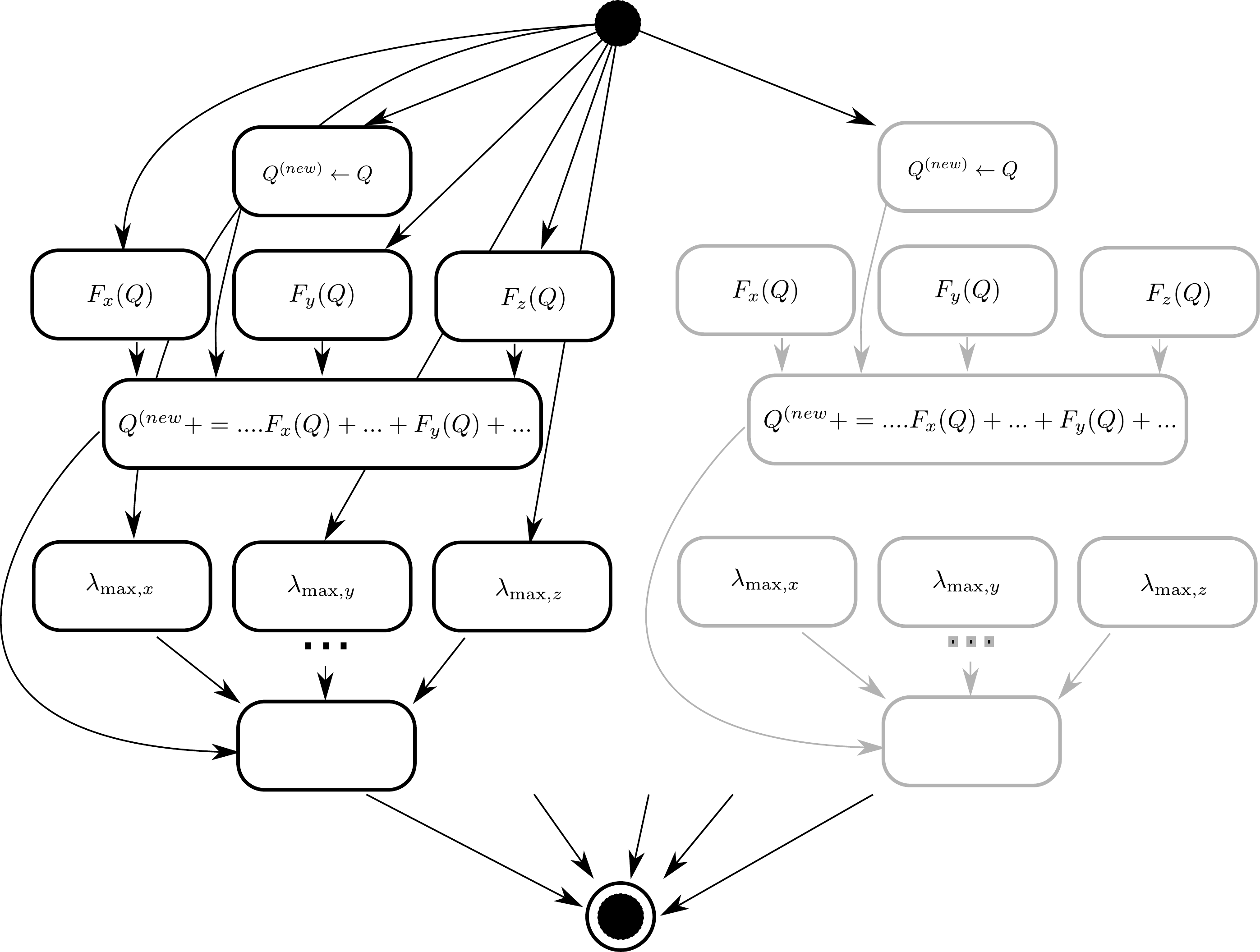}
 \end{center}
 \vspace{-0.4cm}
 \caption{
  Partial sketch of the task graph fed into the task-graph realisation.
  \label{figure:realisation:task-graph-increased-concurrency}
 }
\end{figure}

Different activities (types of microkernels) within the DAG can run in parallel.
An example is the calculations of the directional fluxes $F_x(Q)$, $F_y(Q)$ and
$F_z(Q)$ \replaced[id=R01]{that}{which} have no dependencies.
Both the batched variant and the patch-wise variant serialise these three steps;
the former globally and the latter on a patch basis. 
It is the task-graph approach \replaced[id=R01]{that}{which} does not go down this route and exposes the
concurrency explicitly
(Fig.~\ref{figure:realisation:task-graph-increased-concurrency}).

\section{SYCL implementation}
\label{section:sycl}

Conceptually, mapping either of the three realisations onto SYCL is
straightforward.
There are subtle details to consider however.

%\subsection{Kernels}
\paragraph{Kernels}

The \emph{batched} variant yields a sequence of SYCL \replaced[id=R01]{command
group functions.
They}{compute kernels which} are submitted to a queue and executed in-order. The synchronisation points currently are realised via
\deleted[id=us]{a} waits on the events returned by a kernel launch, i.e.~each
SYCL \replaced[id=us]{submission}{kernel} waits for \replaced[id=us]{the
required predecessors to terminate. Different command groups thus may run in
parallel, but each handles all volumes of all patches subject to one
microkernel only.}{its predecessor to pass on a termination event.} We do not
tie our implementation to an in-order queue, \replaced[id=us]{for which}{though} we could omit the
waits\deleted[id=us]{ for the latter}.

Each \replaced[id=us]{command group}{compute kernel} hosts one large
\texttt{parallel for} \replaced[id=R01]{iterating}{which iterates} over all
patches times all finite \replaced[id=us]{volumes}{volume} within the patches.
Depending on the step type, we might have further embedded loops over the (flux)
direction and the unknowns (for the copy kernel).
We end up with a cascade of $1+d+d=1+2d$ loops (patches, cells along each
direction, flux along each direction)\replaced[id=R01]{. They}{, which} are
collapsed into one large loop.

The \emph{graph-based} variant starts to spawn one SYCL
\replaced[id=us]{command group}{kernel} per patch for the first algorithmic
step.
Each \replaced[id=us]{command group's}{SYCL kernel's} return event is collected
in a vector.
We continue to launch the next set of \replaced[id=us]{command groups}{SYCL
kernels} for the next algorithmic step per patch, augment each
\replaced[id=us]{submission}{kernel invocation} with the corresponding event
dependencies on the previously collected events---where required---and again gather the resulting events.
The dynamic construction of the SYCL DAG resembles the 
batched variant code, where we replace the global waits with individualised
dependencies.
Each individual task within the graph-based variant hosts one \texttt{parallel
for} \replaced[id=R01]{that}{which} traverses a collapsed loop.
The number of loops collapsed is smaller by one compared to the batched
variant.
% , as we do not have to traverse the patches.
% After all, each graph node is tied to one particular patch only.

The \emph{patch-wise} realisation of the kernel requires most attention.
% SYCL does not support nested parallel loops. 
SYCL does not have direct support for nested parallel loops \added[id=us]{or
sequences of loops within one kernel}.
Furthermore, we may only synchronise individual work groups.
Therefore, we map the outer loop onto a \texttt{parallel for} over work groups
via \texttt{nd\_range}.
\deleted[id=us]{Per work group, we may only exploit a fixed number of threads
fired up once:}

\begin{workaround}
 To support nested parallelism, we \replaced[id=us]{manually collapse the
 loops}{ take the union of all iteration ranges
 within the overall compute kernel \emph{for one patch}, multiply this range
 with the number of patches,} and issue one large \texttt{parallel for} over the resulting total loop range \replaced[id=us]{subject to}{as} an
 \texttt{nd\_range} object\deleted[id=us]{which again decomposes the iteration
 range into patch chunks}. 
 \added[id=us]{
  The \texttt{nd\_range} decomposes the iteration range again into
  subranges that are independent of each other, i.e.~correspond to the outer loop.
 }
\end{workaround}

\begin{workaround}
 To support \replaced[id=us]{sequences of parallel loops within one
 kernel}{nested parallelism}, we \deleted[id=us]{take the union of all iteration
 ranges within the overall compute kernel \emph{for one patch} , multiply this
 range with the number of patches, and} issue one \deleted[id=us]{large}
 \texttt{parallel for}\deleted[id=us]{ over the resulting total loop range as an \texttt{nd\_range} which again decomposes the iteration range into patch
 chunks.}
 \added[id=us]{subject to an \texttt{nd\_range}. After each logical loop body,
 we use a workgroup barrier.
 }
\end{workaround}

\begin{workaround}
 To support \replaced[id=us]{sequences of parallel loops with different
 ranges}{nested parallelism}, we take the union of all iteration
 ranges\deleted[id=us]{ within the overall compute kernel \emph{for one patch}
 , multiply this range with the number of patches, and issue one large
 \texttt{parallel for} over the resulting total loop range as an
 \texttt{nd\_range} which again decomposes the iteration range into patch
 chunks}.
 \added[id=us]{
  Per code snippet in-between two barriers, we manually mask out loop indices
  which are not required in this particular step, i.e.~result from the union.
 }
\end{workaround}

\begin{algorithm}[htb]
{\footnotesize
::sycl::range$<4>$ total     \{ T,2+p,2+p,2+p \} \;
::sycl::range$<4>$ workgroup \{ 1,2+p,2+p,2+p \} \;
queue.submit( [\&](::sycl::handler \&handle) \{ \\
\phantom{x}   handle.parallel\_for( \\
\phantom{xx}   ::sycl::nd\_range$<4>$\{total, workgroup\}, \\
\phantom{xx}   [=] (::sycl::nd\_item$<4>$ i) \{ \\
\phantom{xxx}    reconstruct $patch,c$ \;
\phantom{xxx}     if {($c_x\geq 0 \wedge c_x\leq p-1$)}{
                    compute $F_x(Q)(patch,c)$\;
                  }
\phantom{xxx}    i.barrier()\;
\phantom{xxx}    \ldots \ // Next algorithmic step \;
\phantom{xxx}    \ldots \ // with different masking\;
\phantom{xx}  \})\;
\});

}
 \vspace{0.2cm}
  \caption{
    Manual masking \replaced[id=us]{and concatenation of 
    logical algorithmic steps}{within workgroups} for the patch-wise
    realisation.
    \label{algorithm:patch-wise}
 \vspace{-0.4cm}
    }
\end{algorithm}

\noindent
As each patch is mapped onto a workgroup and work groups run concurrently, we
add a workgroup barrier after each compute step on a patch (after each
horizontal cut-through in Fig.~\ref{figure:realisation:task-graph}).
The iteration range per algorithmic step per workgroup is different from step
to step, i.e.~between any two barriers: 
A single flux computation for example runs over an iteration range of
$(2+p)\cdot p^{d-1}$, while the subsequent update of a cell using all
directional fluxes \deleted[id=us]{however only} iterates over a range of
$p^{d}$.
Therefore, the workgroup loops over the maximum iteration range of $(2+p)^d$
right from the start, and we add \replaced[id=us]{\texttt{if} statements}{ifs}
to mask out unreasonable computations manually (Alg.~\ref{algorithm:patch-wise}).

%\subsection{Data structure layout} 
\paragraph{Data structure layout} 

The discussion of AoS vs.~SoA is an all-time classic in supercomputing.
In the context of Finite Volumes, SoA can yield significant speedups.
However, we write \replaced[id=us]{our}{out} kernels around the notion of
microkernels \replaced[id=R01]{wrapping}{which in turn wrap} user functions for the PDEs.
This commitment which allows us to separate the roles of domain scientists
clearly from performance engineers \cite{Gallard:2022:Roles} implies that AoS is
the natural data structure \replaced[id=us]{for}{of} $Q$ in ExaHyPE:
fluxes, eigenvalues, sources, \ldots all are defined as functions over $Q$.
Having all $N$ entries of $Q$ consecutively within the memory allows us to
realise these operations in a cache efficient manner.
Furthermore, many PDEs require the same intermediate results entering all
components of $F_x(Q)$, e.g. 
Storing and traversing $Q$ as SoA would require us either to gather the $N$
entries first, or to recompute partial results redundantly.

For the temporary results, i.e.~all the fluxes and eigenvalues, it is not clear
if they should be stored in AoS or if the microkernels should scatter them
immediately into SoA.
They enter subsequent compute steps where SoA could be beneficial.
Our realisation hence parameterises the microkernels further, such that
we can use them with different storage formats:
We augment the enumerator with a function $enum: [0,\ldots,T-1] \times
[-1,0,\ldots,p]^d \times \mathbb{N}^+_0 \mapsto \mathbb{N}^+_0$, such that they
also take the unknowns into account.

\begin{workaround}
 Due to the absence of $d>3$-dimensional ranges within SYCL, we ``artificially''
 map our higher-dimensional indices and ranges onto $d=3$-dimensional SYCL
 ranges.
\end{workaround}

\noindent
While this is a workaround and higher-dimensional ranges in SYCL are
promised for future language generations, the manual
\replaced[id=us]{partial linearisation}{realisation} allows us to anticipate the
order of the fastest running index within $enum$, such that all memory accesses are coalescent.

\deleted[id=us]{
Our vanilla task graph implementation assembles the task graph
dynamically.
Yet, we also provide a variant with \emph{explicit assembly}, where the task
graph construction is recorded or the task is explicitly constructed on the
host.
Once the whole task graph is assembled, we submit the graph en bloc to the
device.
% \added[id=R1]{
%  Both the dynamic and the explicit assembly happen at runtime for every kernel
%  launch and hence enter the kernel runtime.
}

\paragraph{Reduction\deleted[id=us]{ and data movements}}

Native SYCL reductions for the eigenvalue reduction are available when we work
with the task graph realisation:
We launch one dedicated SYCL \replaced[id=us]{command group with a
reduction}{reduction kernel} per patch.
For the batched variant, we have to abandon the \texttt{parallel\_for} over
a range\deleted[id=R01]{ which we use otherwise}, and instead launch
the for over an \texttt{nd\_range} similar to the patch-wise solution. 
Within the for loop, we use the \texttt{reduce\_over\_group} variant.
The patch-wise variant finally is able to use the reduction over the
group, too.
However, the manual masking has to be mapped onto a branching \replaced[id=R01]{that}{which} returns the
neutral element, i.e.~$\max ( \lambda _{\text{max},x}(Q), \lambda
_{\text{max},y}(Q), \lambda _{\text{max},z}(Q) ) = 0$.

Without \texttt{reduce\_over\_group}, it is possible to realise the reduction via atomics or to run
through each patch serially.
A serial implementation means that the batched flavour launches a
\texttt{parallel for} over the patches (concurrency $T$) yet uses a plain nested
C for loop for the reduction itself.
For the patch-wise implementation it means that only the 0th thread per patch
aka workgroup performs this loop.
All the others are masked out.
\added[id=us]{We use the parallel variant.}

\paragraph{\added[id=us]{Data transfer}}

\added[id=R03]{
SYCL offers USM.
Hence, we can run all kernel variants directly on the GPU,
as long as the underlying data structures are allocated via
\texttt{malloc\_shared}.
The responsibility to transfer all data then is deployed to the runtime.
Alternatively, we can copy data forth and back explicitly.
Remote GPU allocations issued by the host, as required by an explicit copy, are
expensive in OpenMP \cite{Wille:2023:GPUOffloading}.
Assuming similar patterns arise for SYCL, 
we implement managed memory, 
where we allocate memory explicitly on the GPU yet do not free it once a kernel
has terminated.
Instead, we recycle this memory.
}

\added[id=R03]{
 For the batched kernel variant, there is limited freedom to overlap
 memory transfer and kernel kick off.
 The piece-wise and the DAG variant however allow the kernels to
 overlap by kicking off one patch's
 computations while data for the second patch is still dropping in.
 Explicit copying and the managed memory eliminate this advantage, as they are
 realised as preamble to the compute kernel launch, but USM plus patch-wise or
 DAG flavour allow for overlaps of computations and data transfer.
}

\paragraph{\added[id=us]{Code change complexity}}
\added[id=R01]{
 All three realisation flavours have to be maintained separately.
 As we rely on the notion of a microkernel and can extract common features such
 as the enumerations, the allocation of temporary data structures, or the actual
 data transfer (if required) into helper functions, each
 manifests in a few hundred lines of code realising the core SYCL task or loop
 structure.
}

\added[id=R01]{
 The structure of the batched realisation in SYCL resembles 1:1 a plain C++
 implementation subject to the additional queue submits and waits.
 The patch-wise variant is similar to its C++ counterpart, yet requires one
 additional barrier per algorithmic step plus the additional masking (if
 statements).
 This corresponds to two lines of additional code per algorithm step and an
 increased logical complexity to keep track of the correct masking.
 The task-graph version resembles the batched variant but does not wait for any
 outcome:
 The tasks are spawned in the same order as in the batched variant, the
 task completion events are stored in one STD vector per algorithm step, and
 additional \texttt{depends\_on} calls insert the dependency graph's edges
 using the previously populated vectors.
 The task logic---although not trivial---materialises in one or two
 lines of code per algorithmic step.
 All task graph construction is dynamic and hence enters the kernel runtime
 cost. }

\section{Results}
\label{section:results}

Our present work studies exclusively the Euler equations ($N=2+d$) and
\deleted[id=us]{hence} drops the non-conservative terms $\boldsymbol{
\mathcal{B} }_i = 0$ as well as the point ($\delta=0$) and volumetric ($\textbf{S}(Q) = 0$)
sources\added[id=us]{ from (\ref{equation:finite-volumes:PDE})}.
All timings present the compute time for $T$ patches on the GPU, and we
typically use $p \in \{4,6,8\}$.
\deleted[id=R03]{Data transfer cost and compute overhead from ExaHyPE on the
host are eliminated.}
All measurements are averaged over at least \replaced[id=us]{16}{50}
samples\deleted[id=us]{
to avoid statistic outliers}.

We ran our experiments on \deleted[id=us]{one logical segment of} an A100 NVIDIA
GPU\deleted[id=us]{ configured in multi-instance mode}.
The card runs at 1,056 MHz, and it features 80GB of HBM2e memory\deleted[id=us]{
of which we can access 10GB.
That is, we can exploit 14 SMs each hosting 32 CUDA cores for
double precision floating point operations}.
Further to that, we also run all experiments on one
\replaced[id=us]{stack (half of))}{tile} of a \deleted[id=us]{pre-production}
Intel Data Center GPU Max \replaced[id=us]{1550}{1100} (Ponte Vecchio)\deleted[id=us]{.
We use the 300W, 56 X$^e$ core version} equipped with \replaced[id=us]{128}{48}
GB of HBM2e memory and clocked at a base frequency of 1,000 MHz.
\deleted[id=us]{Each of its 56 X$^e$ cores hosts 8 $X^e$ vector engines.
The high-end (600W) PVC variant with
1024 $X^e$ vector engines, or a full A100 or even H100 card likely yield quantitatively
different results.
Also, we also note that the PVC driver and BIOS are still
in active development and hence might improve its performance further.
}

%
% NVIDIA should be able to do 480 flops per cycle
% Intel should be able to do 448 flops per cycle, but the cycles are slower
%

For all experiments, we employed the \replaced[id=us]{2023.2.0 }{2023.1} oneAPI
software stack with Intel's LLVM compiler. 
% Although the icpx compiler did yield slightly better runtime, we only observe this advantage when the floating point precision is reduced.
On NVIDIA, the Codeplay plugin was added to enable the CUDA backend for SYCL,
whereas \added[id=us]{the} Intel hardware is supported out of the box (Level
Zero).
\added[id=us]{
 Our experiments with CUDA unified memory crash immediately due to an error in
 the CUDA layer. 
 On the PVC, the USM realisations pass, although we had to disable
 advanced features (Appendix~\ref{appendix:PVC}).
 We note that both the CUDA software stack and the PVC driver and BIOS are still
 in active development.
 An improved stability might lead to quantitatively different
 results in the future.
}

%
% A paragraph without a subsection, but we need to save space
%
\paragraph{Patch-wise kernels}
\label{section:results:patch-wise-kernels}

\begin{figure}[htb]
 \begin{center}
  \includegraphics[width=0.4\textwidth]{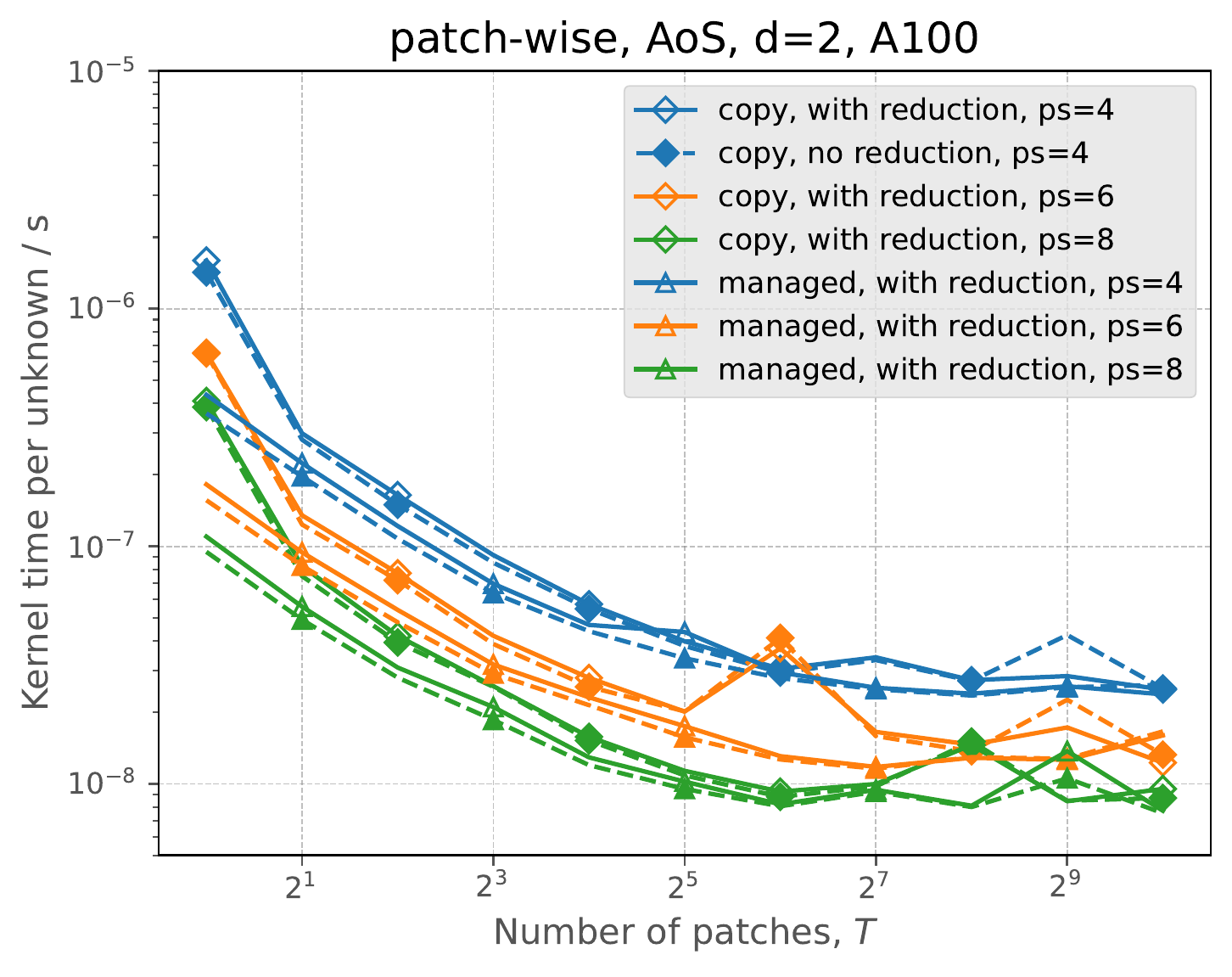}
  \includegraphics[width=0.4\textwidth]{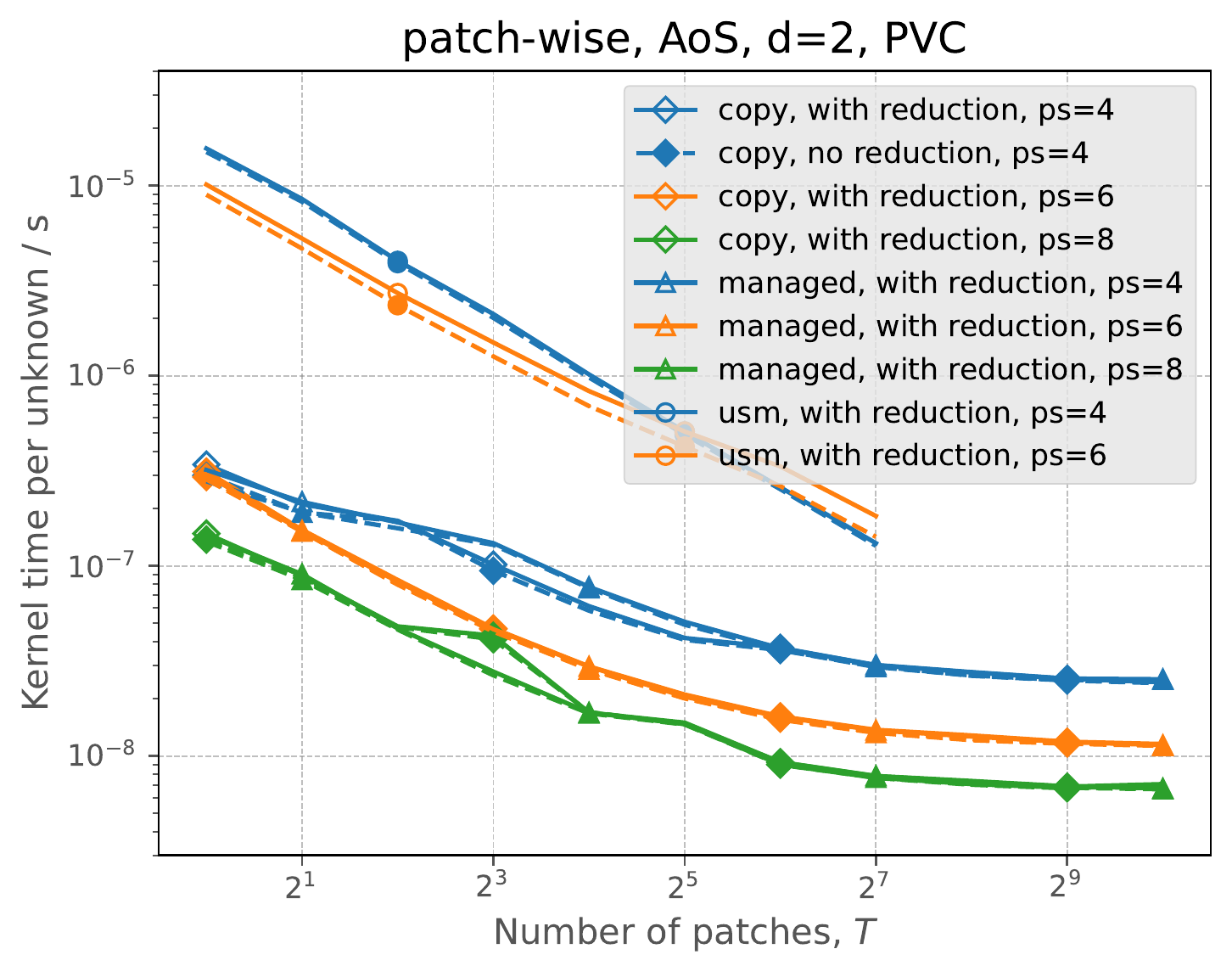}
 \end{center}
 \vspace{-0.6cm}
 \caption{
  Cost per degree of freedom update for various $p$ and $T$
  choices for $d=2$ on an NVIDIA A100 (top) or Intel PVC (bottom). 
  Patch-wise realisation.
  \label{figure:results:2d:AoS:patch-wise}
 \vspace{-0.4cm}
 }
\end{figure}

%
% What we do
%
We start with an assessment of the patch-wise strategy for 
different $p$ values and $T$ choices
(Fig.~\ref{figure:results:2d:AoS:patch-wise}).
AoS is used for all input, output and intermediate data structures.
Our experiments once run the kernel including the computation of a final maximum
eigenvalue, before we rerun the same kernel again yet strip it off this final
reduction. 
\added[id=R03]{
 All timings incorporate data transfer.
}

%
% What we see
%
Higher $p$ or $T$ counts reduce the cost per unknown update\added[id=us]{ on
both architectures and for all realisation flavours.
The reduction increases the runtime, but the impact is almost negligible.
As we increase the problem size, we would expect throughput improvements given
the enormous hardware concurrency of the cards.
Instead, the A100 data plateaus and exhibits deteriorated performance for some
$p \cdot T$ combinations.
The PVC is robust regarding outliers, but plateaus as well.
All $d=3$ data are qualitatively similar, though the anomalies on the A100 are
more pronounced.
Explicit copying and host-managed GPU memory both outperform USM on the PVC,
but do not differ from each other.
On the A100, it is advantageous to
recycle memory.
} 
\deleted[id=us]{. For larger patch sizes, too many patches
make the relative cost per degree of freedom update however plateau, such that smaller
patches at higher $T$-count yield a higher throughput.
Adding reduction brings the
cross-over point forward.
Reruns on the PVC yield qualitatively similar results.
Yet, the reductions have a less severe impact on the runtime;
occasional inversions of the performance curves are due to runtime fluctuations.
The chip perform better for smaller workloads, but is outperformed by the
A100 for bigger $p \cdot T$-combinations. 
For $p=8$, the PVC shows stagnating throughput, independent of $p$ and $T$.
There is no qualitative difference between $d=2$ and $d=3$ runs (not shown),
although the plateau is brought significantly forward and the full stagnation
is observed for $p=6$.
}

\begin{figure}[htb]
 \begin{center}
  \includegraphics[width=0.4\textwidth]{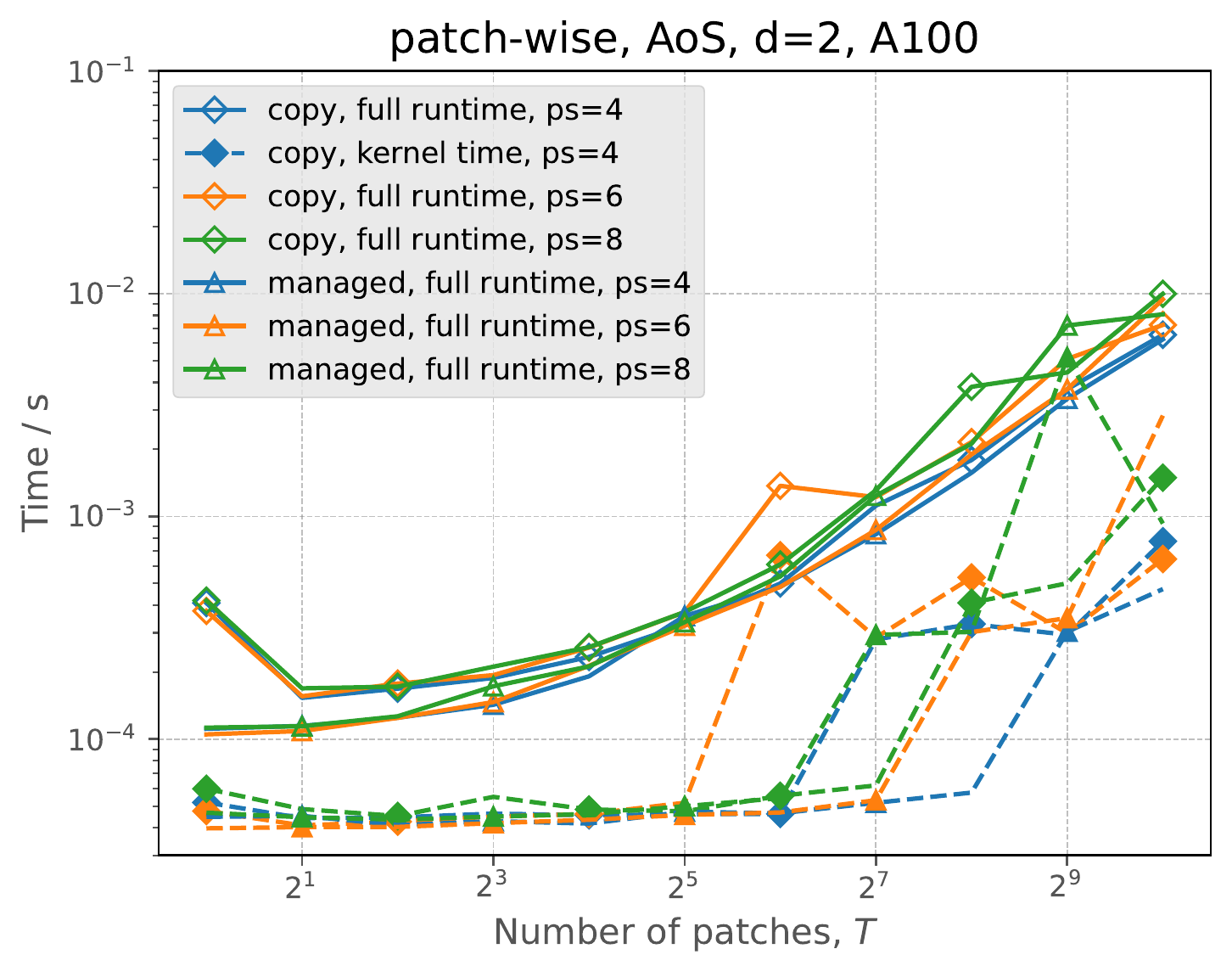}
  \includegraphics[width=0.4\textwidth]{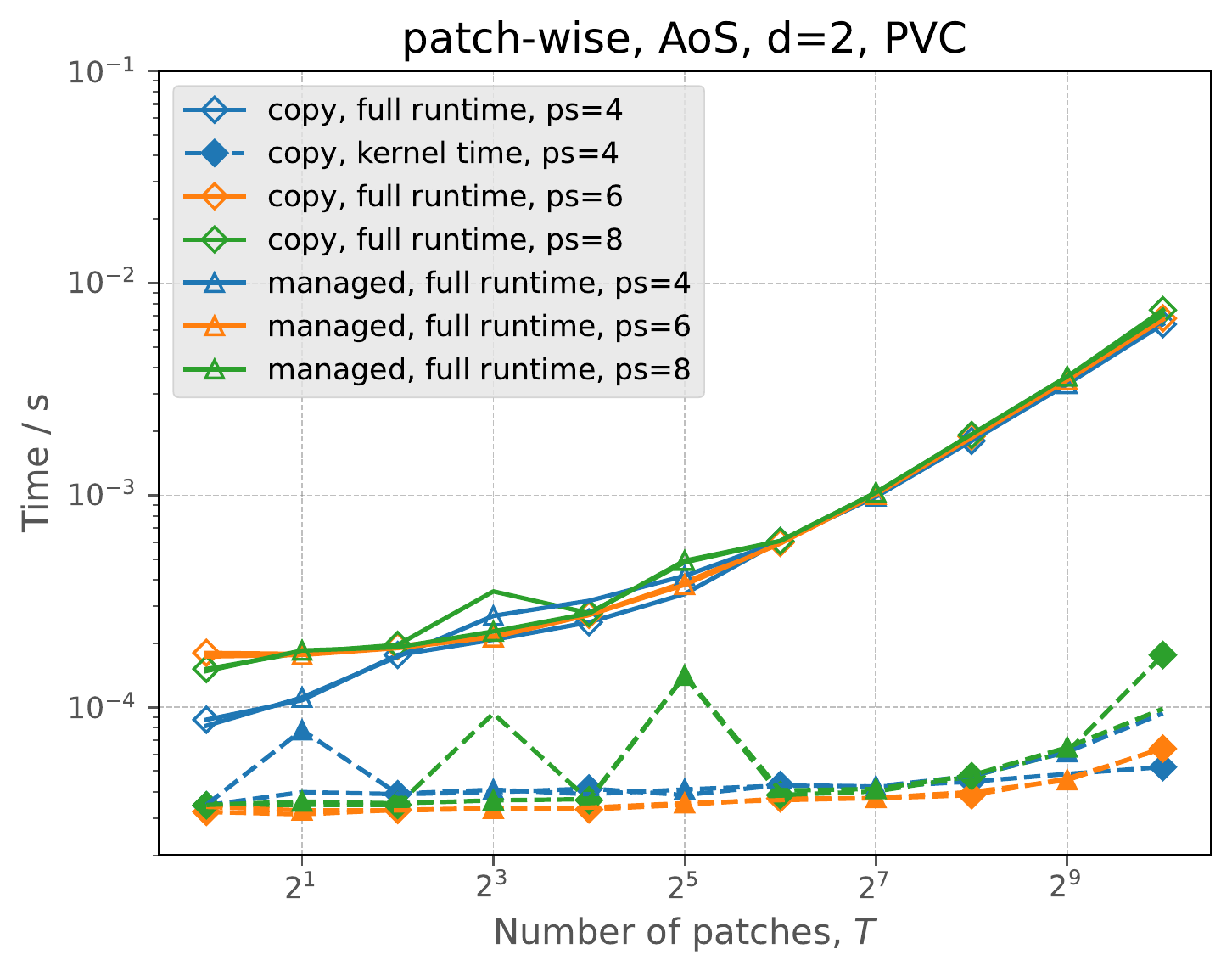}
 \end{center}
 \vspace{-0.6cm}
 \caption{
  \added[id=R03]{
   Breakdown of the total runtime for all patches from
   Fig.~\ref{figure:results:2d:AoS:patch-wise} into total kernel compute time and total kernel runtime including data transfer
   cost.
  }
  \label{figure:results:2d:AoS:patch-wise-break-down}
 \vspace{-0.4cm}
 }
\end{figure}

\added[id=R03]{
 Our runtimes are dominated by data transfers
 (Fig.~\ref{figure:results:2d:AoS:patch-wise-break-down}).
 ExaHyPE's patches are scattered over the CPU's heap memory and have to be
 brought to the accelerator one by one.
 The transfer cost suffers from the latency of multiple, scattered memory
 transfers.
 It therefore increases with $T$.
 Copying a single patch seems to be particularly expensive on the
 A100.
 Here, we benefit particularly from a managed memory approach.
 The compute time flatlines for small $T$.
 Once we increase $T$ sufficiently, also the compute time starts to
 grow.
 The PVC compute times peak for some small $T$ choices and $p=8$, while the A100
 shows some runtime peaks for few larger $p \cdot T$ combinations.
 The latter explain the runtime ``anomalies'' in the total execution
 time.
}

%
% Interpretation and explanation
%
Each patch is mapped onto a workgroup of its own.
Increasing the workload $p$ per
workgroup yields higher throughput, since the elementary workgroup workload is
higher, i.e.~we can do more calculations before we terminate or swap a
workgroup.
Furthermore, the \replaced[id=us]{share}{ratio} of iteration range indices that
do not fit to the hardware's workgroup size and hence have to be masked out diminishes.
The thread divergence decreases.
\deleted[id=us]{
We note that steps like the flux updates in one direction require
$(p+2)\cdot p^{d-1}$ microkernel invocations\replaced[id=R01]{. This}{,
which} explains why $p=8$ for example misfits the PVC architecture.
}
\added[id=us]{
 The PVC seems to be particular sensitive to these effects for small $p$.
 The A100 is sensitive to these effects when they are scaled up by using many
 patches $T$.
 The compute performance's inital flatlining demonstrates that small choices
 of $T \cdot p$ struggle to exhaust the compute power.
}

\begin{workaround}
 Larger $d \cdot p$\replaced[id=us]{ combinations}{-values} are infeasible
 \deleted[id=us]{for $d=3$} due to workgroup size limits in SYCL (1,024 on both PVC and A100).
 Large patches $p$ have to broken down manually such that they fit onto the
 hardware.
\end{workaround}

\deleted[id=us]{
\noindent
As keeping many large workgroups with multiple
subworkgroups (warps) in flight becomes expensive anyway---our profiling suggests that we have to blame register spilling---it is
advantageous to constrain the $p$ choices right from the start.
This is a counterintuitive finding:
Common knowledge suggests that large work sets are always advantageous on GPUs.
}
\added[id=R03]{
 \noindent
 We also conclude that USM---even ignoring the reported stability issues---is
 best to be avoided if we can copy explicitly.
 Notably, it fails to to benefit from overlapping computations and calculations.
 However, our implementation lacks the usage of prefetches which might help to
 make the USM implementation faster.
%  The latter otherwise performs well and the data transfer cost curve is smooth
%  without any peaks.
%  If we have multiple threads creating GPU patches and if these patches become
%  available at different points in time, it pays off to use the managed memory
%  allocation, i.e.~to reserve a chunk of memory once and then to hand this one out rather
%  than repeated device allocations.
%  If the GPU-ready patches all become available around the same time, it is best
%  to assemble them into one batch, aka meta-patch, and to deploy them to the GPU
%  en bloc.
}

% This confirms similar observations with OpenMP and therefore is likely a
% hardware property \cite{Wille:2023:GPUOffloading}.
%work cannot saturate the GPU for

%
% A paragraph without a subsection, but we need to save space
%
\paragraph{Batched kernels}

\begin{figure}[htb]
 \begin{center}
  \includegraphics[width=0.4\textwidth]{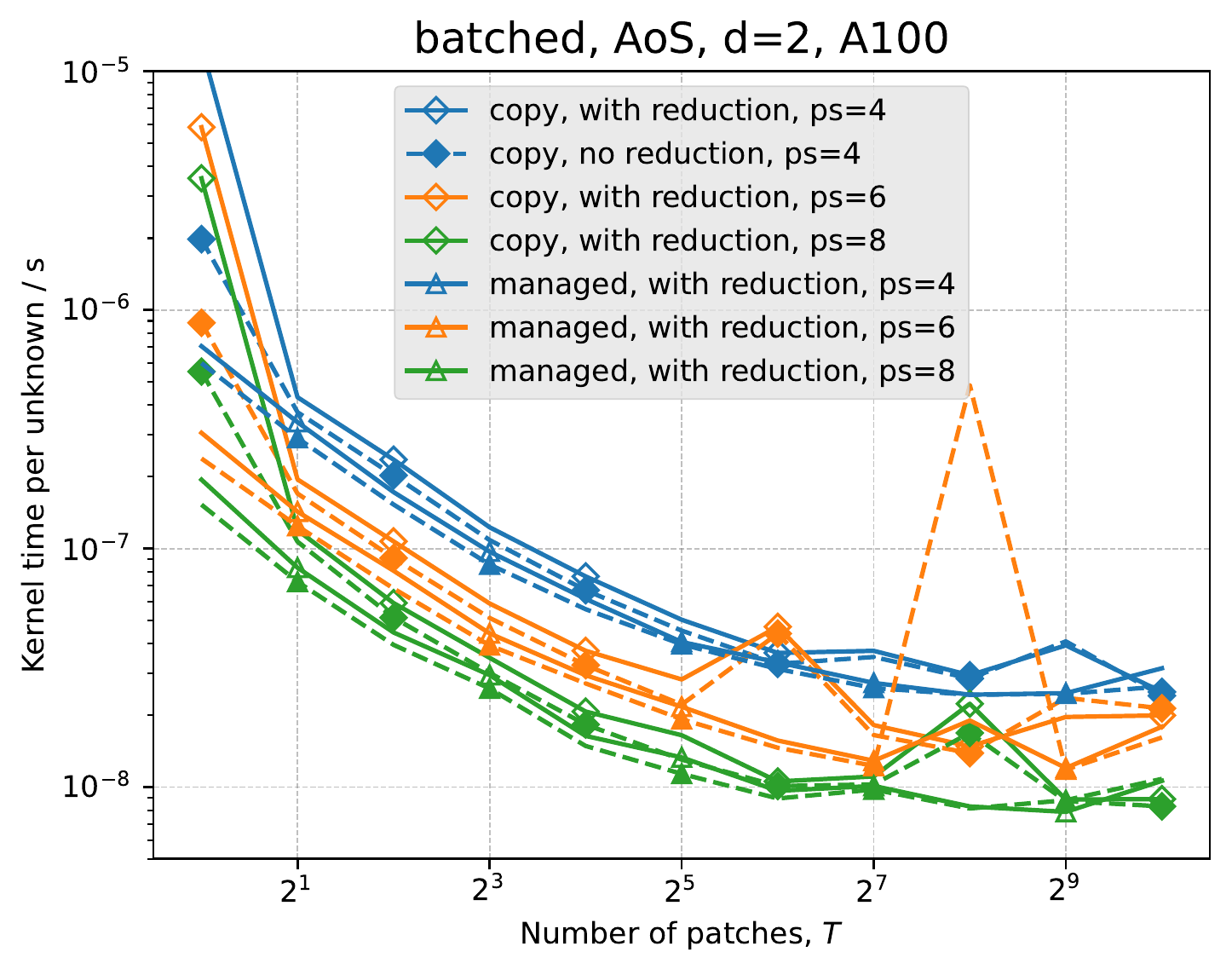}
  \includegraphics[width=0.4\textwidth]{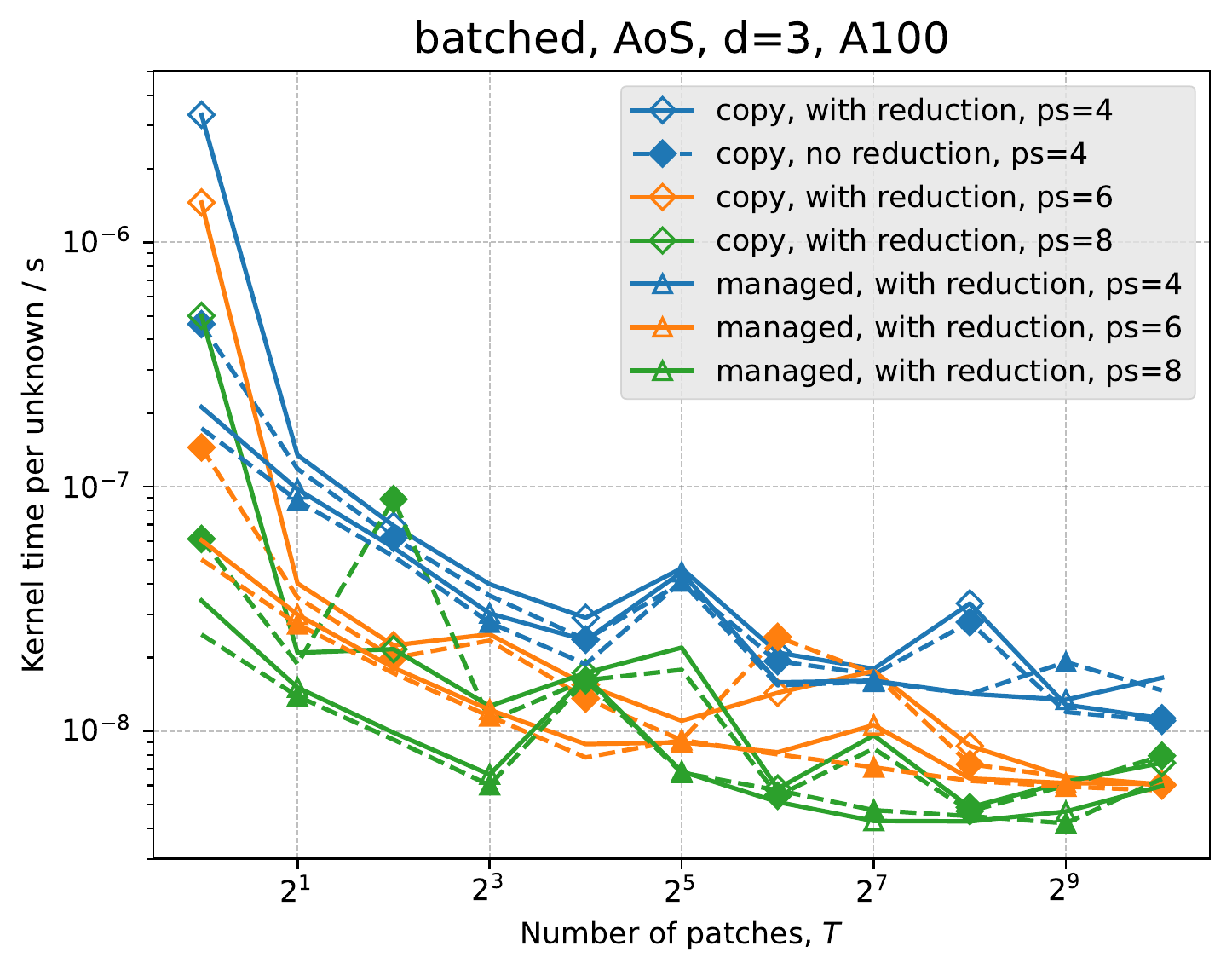}
 \end{center}
 \vspace{-0.6cm}
 \caption{
  Normalised runtime for batched kernels on the A100 for
  \replaced[id=us]{$d=2$ (top) and $d=3$ (bottom)}{$d=3$}.
  \label{figure:results:A100:batched}
 }
\end{figure}

\begin{figure}[htb]
 \begin{center}
  \includegraphics[width=0.4\textwidth]{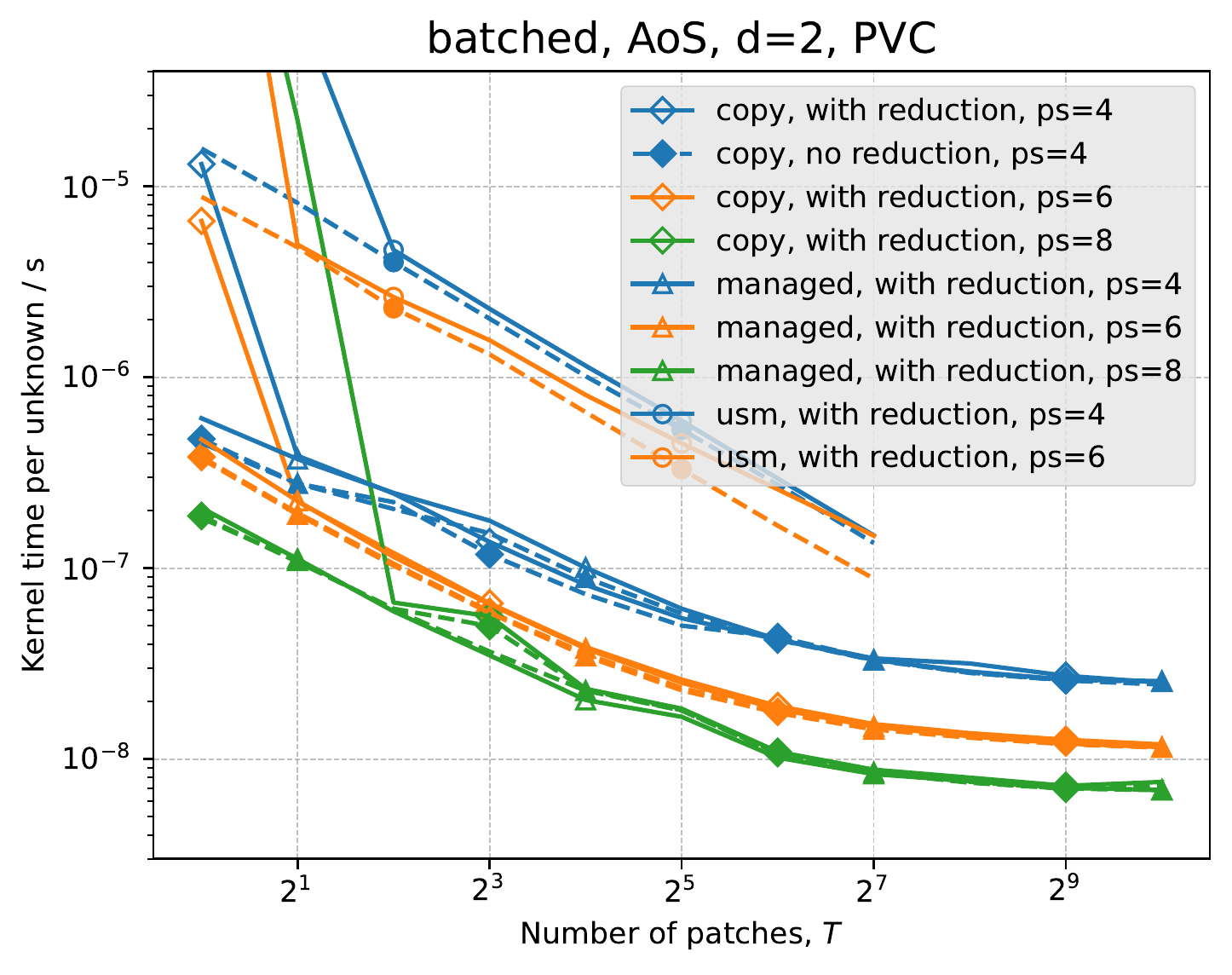}
  \includegraphics[width=0.4\textwidth]{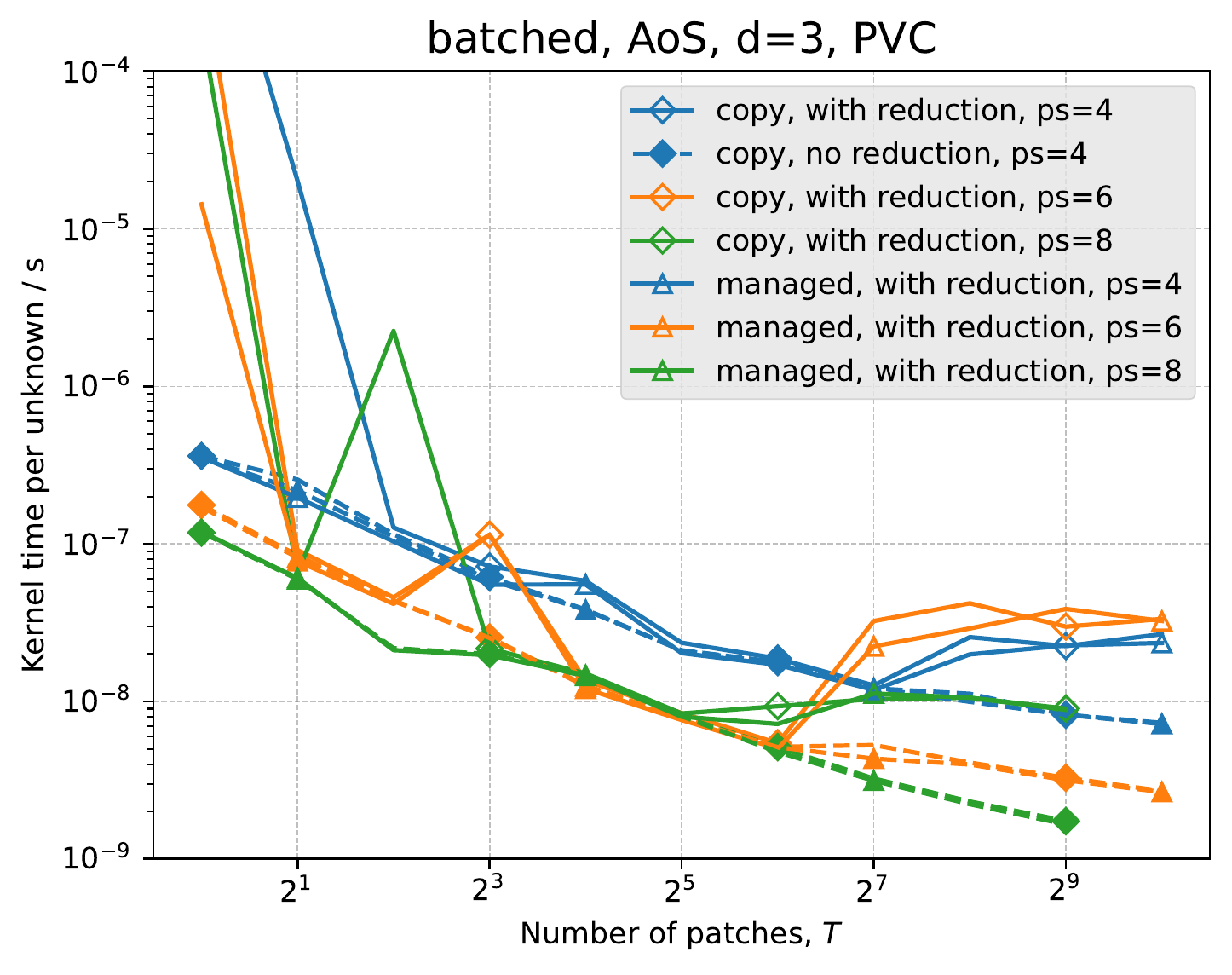}
 \end{center}
 \vspace{-0.6cm}
 \caption{
  Normalised runtime for batched kernels on PVC for $d=2$ (top), and on PVC
  for $d=3$ (bottom).
  \added[id=us]{
   Deteriorated runtimes $t(T=1)\gg 4\cdot10^{-5}$ are cut off to preserve a
   uniform scaling of the y-axes.
   USM data is omitted due to its non-competitiveness.
  }
  \label{figure:results:PVC:batched}
  \vspace{-0.4cm}
 }
\end{figure}

%
% What do we see
%
The batched kernel variant \replaced[id=us]{yields performance
curves on the A100 where the peaks are amplified. It also is significantly
slower than the patch-wise approach for a single patch (Fig.~\ref{figure:results:A100:batched}).
The PVC's performance
deteriorates totally for $T=1$ and for $d=3$ with larger $T \cdot p$
(Fig.~\ref{figure:results:PVC:batched}).
Here, we suffer from the cost of the reduction.
For all other experiments, 
the batched variant 
catches up with its patch-wise cousin and eventually matches its performance
on either machine as $T \cdot p$ grows.
}{eliminates the cross-over effects between different $p$
curves besides for one PVC data point for $d=3,p=4$ (Figs.~\ref{figure:results:A100:batched} and \ref{figure:results:PVC:batched}).
While it otherwise delivers monotonously decreasing cost as $T$ or $p$ increase,
the A100 curves are close to constants.
On the PVC, we see kernels with reduction sometimes outperforming their
counterparts without a reduction.
}

%
% Interpretation
%
The GPU copes well with a large number of threads \replaced[id=R01]{that}{which} all run the same
computations, as SYCL can subdivide the iteration range into workgroups
as fit for purpose.
\replaced[id=us]{Per algorithmic step, the}{
The} batched variant has no logical thread divergence, i.e.~we do not mask out
threads manually, although SYCL will add idle threads internally to make the
iteration ranges match the hardware concurrency.
A profiler indeed shows smaller workgroup sizes compared to
the patch-wise version, i.e.~more workgroups are created\replaced[id=us]{. Yet,}{
but} the register pressure per workgroup is significantly smaller.
\deleted[id=us]{
 For some $T,p$ combinations, this
 splitting seems to be particularly advantageous on the PVC.
}

\replaced[id=us]{
 The ratio of idling threads per workgroup compared to the actual workload is
 smaller than for the patch-wise variant, and the arrangement of workgroups
 might even change between different algorithmic kernel steps.
 However, we pay a price for launching
 multiple GPU kernels in a row and the corresponding synchronisation points.
 The A100 seems to be particularly sensitive to this.
 $p \cdot T$ has to become reasonably large, before the increased
 flexibility plus the fewer ``wasted'' threads allow the batched version to compensate for the kernel
 launch overhead and to close up on the patch-wise realisation.
 While we do not have to care about maximum workgroup sizes for the kernel's
 main compute steps, an efficient, parallel reduction of the final maximum
 eigenvalue continues to hinge on the fact if one patches fits into a workgroup.
 Even if this is the case, the PVC's performance might suffer from
 reductions over large data sets.
}{ For some $T,p$ combinations, this
 splitting seems to be particularly advantageous on the PVC.
}

\deleted[id=us]{
Across the board, using larger patch counts and larger per-patch sizes pay
off. 
Since the batched version submits a sequence of kernels with strict
dependencies, it however introduces synchronisation into the realisation.
On both A100 and PVC, the batched variant hence ends up being at least
a factor of two more expensive than its patch-wise counterpart.
}

%
% A paragraph without a subsection, but we need to save space
%
\paragraph{Task-graph kernels}

\begin{figure}[htb]
 \begin{center}
  \includegraphics[width=0.4\textwidth]{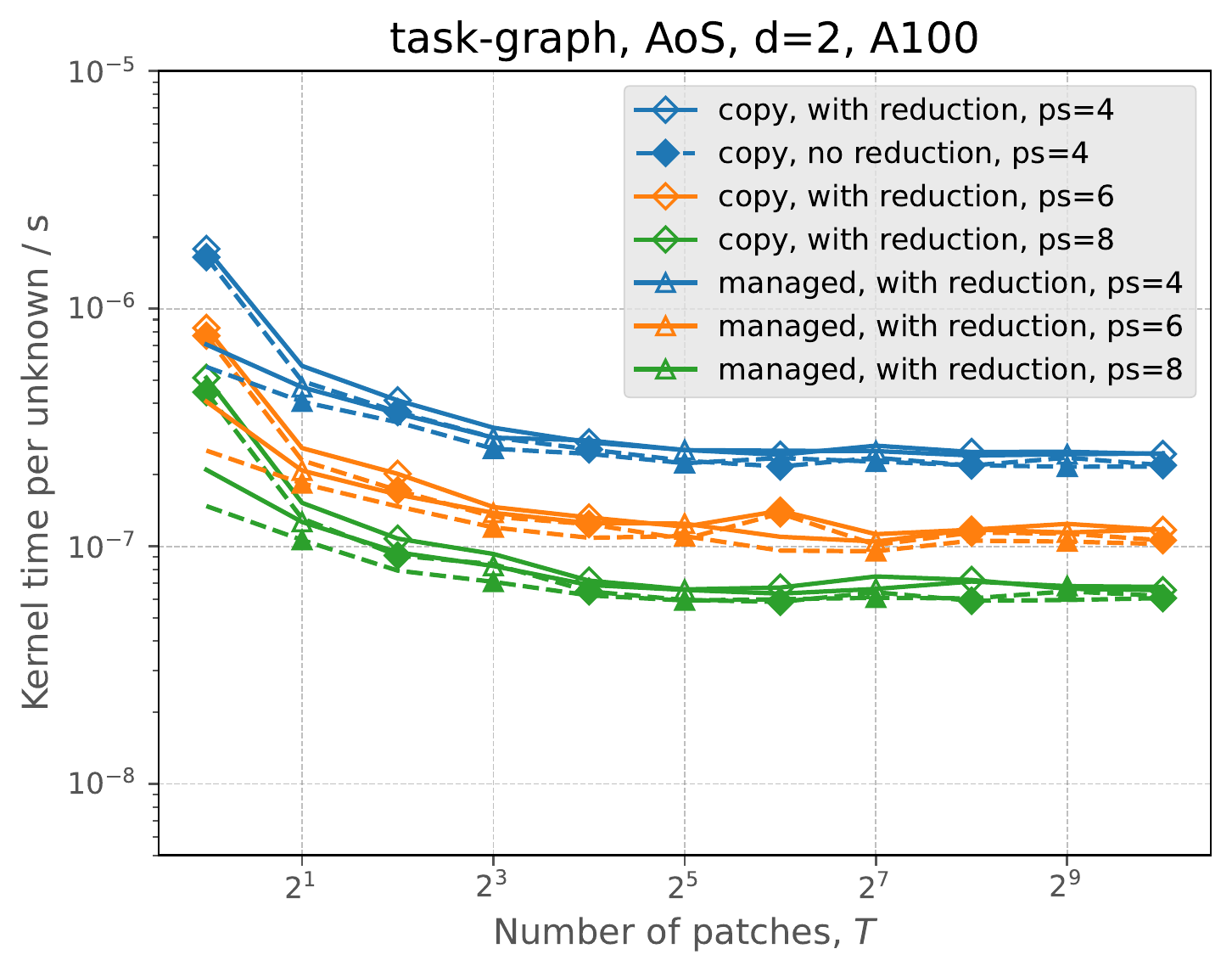}
  \includegraphics[width=0.4\textwidth]{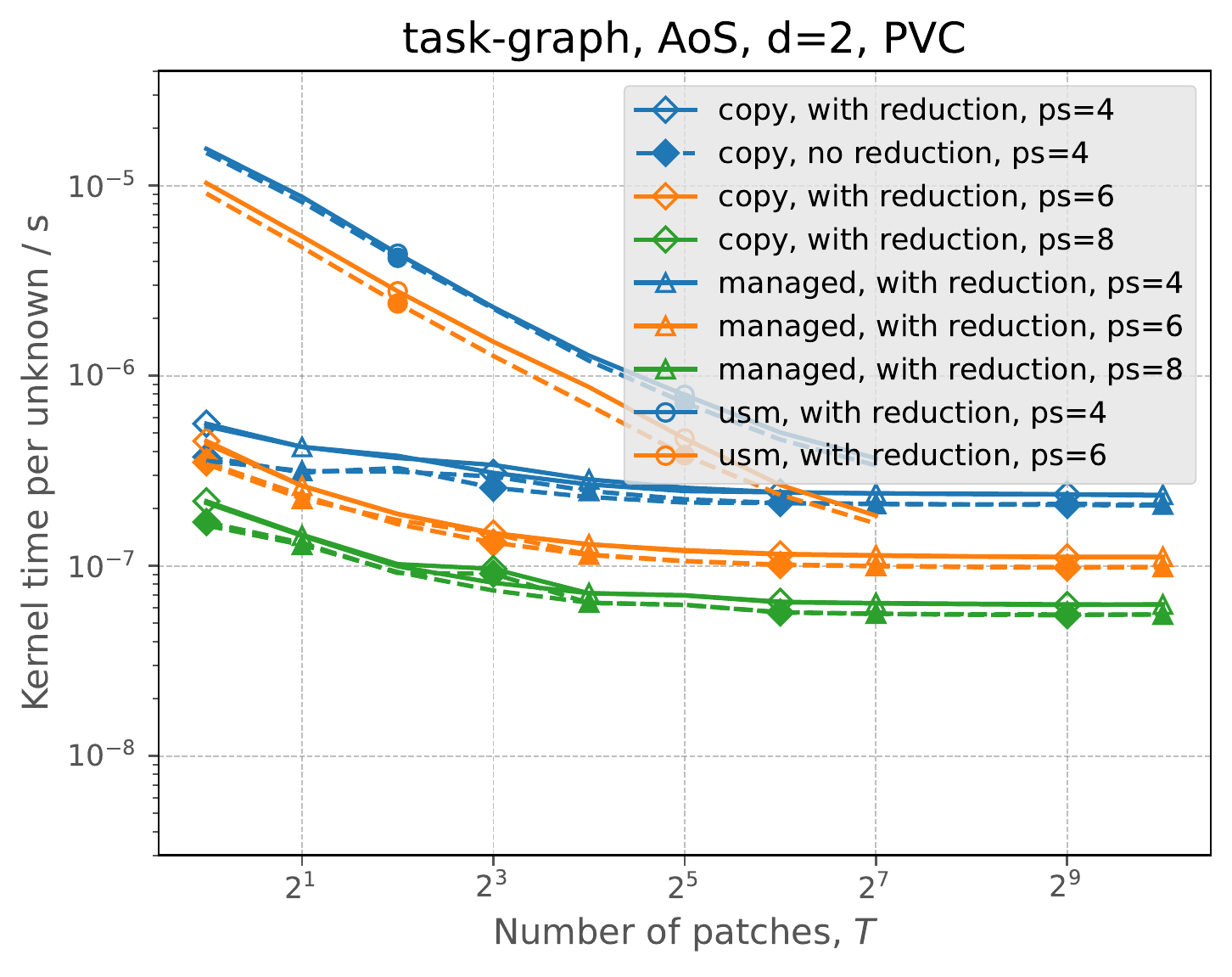}
 \end{center}
 \vspace{-0.6cm}
 \caption{
  Kernel runtimes on the A100 \added[id=us]{(top) and the PVC (bottom)} for a
  realisation using a SYCL task graph for $d=2$.
  \label{figure:results:PVD:task-graph}
 }
\end{figure}

%
% What we see
%
A straightforward implementation of our task graph approach does
not scale at all in $T$ (Fig.~\ref{figure:results:PVD:task-graph}).
We observe qualitatively similar data for $d=2$ and $d=3$ on
\replaced[id=us]{the A100}{both test systems}.
While increasing $p$ brings the relative runtime down, the
\added[id=us]{performance immediately starts to stagnate and the}
resulting overall kernel remains significantly slower than the batched or patch-wise kernels.
\added[id=us]{
 On the PVC, the performance does not stagnate but actually increases for
 growing $T$ after a brief scaling phase or the code deadlocks (no data shown).
}
\added[id=R03]{
 For $d=2$, the performance gap between USM and the other approaches
 eventually closes, but this is due to the poor performance of the latter rather
 than USM's performance.
}

%
% Why is that
%
We assume that the dynamic assembly of the task graph\added[id=R01]{, which is
included in the compute kernel timings,} is too expensive, as it maps each
compute step within the DAG onto a separate kernel launch.
\deleted[id=us]{Profiling shows the GPU idling/stopping in-between two
launches.}
The low arithmetic intensity of our individual computations amplifies this
effect.
Once $p$ increases and hence introduces reasonable high compute load per DAG
step, the relative overhead reduces.
\added[id=us]{
 The A100 profits from this fact, while the PVC continues to struggle with the
 high number of kernel launches.
}
We conclude that a dynamic administration of the task graph on the host side is
not competitive with our two alternative realisation variants.

%
% What to do
%
While our code base should support dynamic numbers of $T$---we want to support
dynamic AMR where we do not know a priori how many tasks are spawned and
offloaded to the GPU per thread
\cite{Li:2022:TaskFusion,Wille:2023:GPUOffloading}---the task graph is fixed
once we decide to load a certain number $T$ of patches with a given $p$ to the
accelerator.
There is no need to assemble a task graph dynamically.
Instead, we can precompile the graph, offload it to the GPU, and let the GPU
handle the dependency administration internally.
SYCL provides (experimental) extensions for this through task graph recording
and explicit graph construction APIs.
Unfortunately, we were not able to use them successfully with our current
software stack.
\added[id=R01]{
Instead, we use dynamic task graph assembly and let the 
assembly contribute towards the kernel runtime.
}

%
% A paragraph without a subsection, but we need to save space
%
\paragraph{Further implementation remarks}

We did extensive studies comparing SoA against AoS and also AoSoA, where the
data data per patch are stored as SoA, while the individual patch data chunks are
stored one after another.
These ordering variations make no significant difference to the runtimes.
Solely converting the input $Q$ into SoA increases the runtime dramatically, as
we then have to gather data for each and every microkernel.

We use SYCL's \texttt{reduce\_over\_group} for reductions wherever a plain
reduction does not work.
It yields a speedup of around a factor of two
compared to a purely sequential version for small patches, and still a runtime improvement of around 10\% for
the larger $p$ values.
An alternative parallel reduction using one atomic per patch is not competitive.
\added[id=us]{
 No problem was scaled up to a point where SYCL's workgroup size becomes a
 limiting factor, though this limitation is discussed.
}

In all kernels, we refrain from an explicit parallelisation over the unknowns
in $Q \in \mathbb{R}^N$.
Some loops over microkernels expose concurrency in the unknowns; to update all components
according to an explicit Euler, e.g.
However, exploiting this concurrency seems never to pay off, while it
\deleted[id=us]{even} introduces massive (logical) thread divergence penalties
in the patch-wise variant\deleted[id=us]{ if we
have to spawn one thread per unknown yet work on the struct level most of the time}.

\section{Outlook and conclusion}
\label{section:conclusion}

SYCL is still the new kid on the block when it comes to GPU programming.
However, its ``all-in-native-C++''--policy makes it attractive to
\replaced[id=us]{projects that}{programmers who} aim for code which runs both on
CPUs and GPUs \cite{Deakin:2022:HeterogeneousProgramming}, and
\replaced[id=us]{that aim to hide platform specifics by sticking to one
language and one implementation only}{performance engineers who have to
interact closely with domain and numerics specialists familiar with C only}.
It also is appealing, as it offers a task-first approach to heterogeneous
programming:
All kernel submissions can be considered to be tasks, return events, and can
have dependencies.
% There is even the opportunity to work with specialised host tasks and have
% inter-queue dependencies.

%
% Flexibility does not pay off
%
Unfortunately, \replaced[id=us]{the}{this} flexibility
\added[id=us]{promised by tasks} seems not to pay off\deleted[id=us]{
universally}.
Our data suggest that \replaced[id=R02]{
 performance engineers have to embed the logical task graph into sequences
 of nested loops and map these loops onto one high-dimensional SYCL loop to
 obtain high performance.
 This is a low-level rewrite of the high-level task concept.
 We discuss two low-level realisations: batched and patch-wise.
 The patch-wise
 variant's performance is robust for small problem sizes and very close to
 how domain scientists traditionally phrase their algorithms (run over all
 patches; per patch, compute \ldots), but it requires some significant re-engineering
 and several workarounds if we implement it in SYCL.
 It is also subject
 to hardware (workgroup size) constraints.
 The batched version outperforms its
 patch-wise cousin for some bigger setups on some machine--dimension
 combinations.
}

\added[id=us]{
 Eventually, performance engineers might have to maintain two
 realisations to facilitate high throughput on a GPU:
 A patch-wise version for small $T$ counts, and a batched version for bespoke,
 larger setups; which are time-consuming by definition.
 On the CPU, a task-based version might be advantageous, which adds a third
 implementation variant.
}
It will be subject of future work to investigate if this pattern changes with
new hardware and software generations, i.e.~if SYCL's native task
formalism catches up performance-wisely.
\added[id=R02]{
As it stands, maintaining different realisations of one and the same numerical
kernels remains necessary though being tedious and error-prone.
}

A fast, native, task-first \added[id=us]{USM} programming paradigm for GPUs
would be \replaced[id=us]{another}{a} real selling point \replaced[id=us]{for
SYCL.
As it stands, platform-portability might be guaranteed for the most generic and flexible way
to phrase computations, but performance 
sacrifices have to be made if we want to use a strict task formalism and do not
want to administer or copy data ourselves.
}{behind
SYCL due to the opportunity to express popular numerical patterns directly within the language.}

%
% Missing features or abstraction level right?
%
\added[id=us]{On the low-level realisation side,}
SYCL offers a GPU abstraction somewhere halfway in-between OpenMP and CUDA:
Some details such as the orchestration of collapsed loops are hidden
(like in OpenMP), while they can be exposed and tailored towards the machinery
explicitly by using \texttt{nd\_ranges}. 
Our work showcases that some codes would benefit from an even higher level of
abstraction.
It is not clear why SYCL does not provide stronger support for
nested parallel fors as we get them natively in OpenMP, 
it is not clear why the
maximum workgroup size \replaced[id=us]{of}{on} the hardware imposes constraints
on the implementation and cannot be mitigated within the SYCL software layer,
and a better support for nested parallelism and the automatic serialisation over
some iteration indices (i.e.~of loops over unknowns) would streamline
the development.

%
% DSLs, future work
%
\replaced[id=R03]{
 While our work makes statements on efficient compute kernel realisations, it
 falls short of examining the impact of further advanced realisation
 techniques such as memory prefetching and task graph recording plus re-usage.
 It might be possible that such invasive techniques---in the
 sense that more manual source code augmentation becomes necessary---allow us to
 close the performance gap between USM and manual memory movements as well as
 on-the-fly task graph construction and task graph processing.
 Further to studying these features, future}
{Future} work will comprise data layout optimisations beyond
simple reordering of temporary data and the optimisation of the core
calculations.
As long as we stick to the policy that no user code is altered, our code is
inherently tied to AoS and the physics calculations are locked away from further
tuning.
If we want to keep the strict separation of roles, it hence might become
necessary to switch to domain-specific languages, such that
the translator has the opportunity to alter both the kernel and the
microkernels.

\section*{Acknowledgements}

Our work has been supported by the UK's ExCALIBUR
programme through its cross-cutting project EX20-9 \textit{Exposing Parallelism: Task Parallelism}
(Grant ESA 10 CDEL) made by the Met Office and the EPSRC DDWG projects
\textit{PAX--HPC} (Gant EP/W026775/1) and 
\textit{An ExCALIBUR Multigrid Solver Toolbox for ExaHyPE} (EP/X019497/1).
Particular thanks are due to Intel's Academic Centre of
Excellence at Durham University.
This work has made use of the Durham's Department of Computer Science NCC
cluster.
Development relied on 
the DiRAC@Durham facility managed by the Institute for Computational Cosmology
on behalf of the STFC DiRAC HPC Facility
(\href{www.dirac.ac.uk}{www.dirac.ac.uk}). The equipment was funded by BEIS capital funding via STFC capital grants ST/K00042X/1, ST/P002293/1, ST/R002371/1 and ST/S002502/1, Durham University and STFC operations grant ST/R000832/1. DiRAC is part of the National e-Infrastructure.

The authors wish to thank 
Andrew Mallinson (Intel) for establishing the collaboration with Intel, 
Dominic E.~Charrier (AMD) for the
initial suggestion to rewrite the OpenMP code with microkernels,
Mario Wille (TUM) for the many discussions around the OpenMP compute kernels,
and all the colleagues at Codeplay and Intel (notably Heinrich Bockhorst for
his help and for reproducing all experimental steps on in-house hardware) for
their help\deleted[id=us]{ and constructive remarks}.

\bibliographystyle{plain}
\bibliography{paper}

\newpage
\cleardoublepage
\appendix

\setcounter{algocf}{0}

\section{Download and build}
\label{appendix:download-and-compilation}

All of our code is hosted in a public git repository 
on \url{https://gitlab.lrz.de/hpcsoftware/Peano} and available to clone. 
\replaced[id=R01]{
  Our GPU benchmark scripts are merged into the repository's
  main, i.e.~all results can be reproduced with main branch. 
  Yet, to use the exact same code version as used for this paper, please switch to 
  tag \texttt{a3461cb6} on the \texttt{gpu} branch.
}{
For this paper, we created the \texttt{gpus-sycl-siam} branch to reproduce the presented results. 
}

\begin{bash}[htb]
% {\footnotesize
git clone -b gpus https://gitlab.lrz.de/hpcsoftware/Peano\;
cd Peano\;
libtoolize; aclocal; autoconf; autoheader; cp src/config.h.in .\;
automake {-}{-}add-missing\;
% }
  \vspace{0.2cm}
  \caption{
    Cloning the repository and setting up the autotools environment. 
    \label{code-snippet:autokernels}
    }
\end{bash}

\begin{bash}[htb]
./configure CC=icx
CXX=icpx LIBS="-ltbb" LDFLAGS="-fsycl -fsycl-targets=nvptx64-nvidia-cuda
-Xsycl-target-backend=nvptx64-nvidia-cuda {-}{-}cuda-gpu-arch=sm\_80"
CXXFLAGS="-O3 -std=c++20 -fsycl -fsycl-targets=nvptx64-nvidia-cuda
-Xsycl-target-backend=nvptx64-nvidia-cuda {-}{-}cuda-gpu-arch=sm\_80" {-}{-}with-multithreading=tbb {-}{-}enable-exahype {-}{-}enable-blockstructured {-}{-}enable-loadbalancing {-}{-}with-gpu=sycl\;

./configure CC=icx
CXX=icpx LIBS="-ltbb" LDFLAGS="-fsycl" CXXFLAGS="-O3
-std=c++20 -fsycl" {-}{-}with-multithreading=tbb {-}{-}enable-exahype {-}{-}enable-blockstructured {-}{-}enable-loadbalancing {-}{-}with-gpu=sycl\;
  \vspace{0.2cm}
  \caption{
    Configure command used for our A100 tests (top) and for the PVC runs
    (middle). \deleted[id=us]{The bottom command is to be used with the Intel toolchain
    (\texttt{icpx}).}
    \label{code-snippet:configure}
    }
\end{bash}

While the project supports CMake as discussed in the project documentation
(available by running \texttt{doxygen documentation/Doxyfile}), we present the
setup using autotools here (Commands~\ref{code-snippet:autokernels}).
To create the actual makefiles for A100 tests, initialise the oneAPI environment
\deleted[id=us]{that comes with the LLVM compiler} or use your native oneAPI modules, and
afterwards configure your code accordingly (Commands~\ref{code-snippet:configure}).
You will obtain a plain makefile \replaced[id=R01]{that}{which} builds all of Peano's and ExaHyPE's core
libraries.

\section{Execution and postprocessing}
\label{appendix:execution-and-postprocessing}

All experiments as discussed are available through a Python script \replaced[id=R01]{that}{which} links a
test case driver (miniapp) to Peano's and ExaHyPE's core libraries.
This miniapp sweeps through the parameter combinations of interest for the
present discussions.

\begin{bash}[htb]
cd benchmarks/exahype2/euler/kernel-benchmarks\;
python kernel-benchmarks-fv-rusanov.py {-}{-}dim 3 {-}{-}patch-size 8\;
  \vspace{0.2cm}
  \caption{
    Build a $d=3$ setup
    with patches of size $8 \times 8 \times 8$.
    The Python script internally invokes \texttt{make} (or \texttt{cmake} if
    you have previously configured the code with CMake) and yields an executable
    \texttt{./kernel-benchmark-fv-Xd-patch-size-Y} where \texttt{X} is the
    dimension (3 in this example) and $Y$ the patch size (8).
    \label{code-snippet:build}
    }
\end{bash}

To build the miniapp, we change into Euler's
\texttt{kernel-benchmarks} folder. 
The Python 3 script \texttt{kernel-benchmarks-fv-rusanov.py} yields the
test case executable (Commands~\ref{code-snippet:build}).
Passing \texttt{-h} provides instructions on various further options. 
For example, the test can be asked to validate the GPU outputs for
correctness against a CPU run, or you can alter the number of samples taken,
i.e.~over how many runs the measurements should average.

\begin{bash}[htb]
export ONEAPI\_DEVICE\_SELECTOR=level\_zero:0\;
for i in 4 6 8; do\; 
    python kernel-benchmarks-fv-rusanov.py -d 3 -ps \$i\;
    ./kernel-benchmarks-fv-rusanov-3d-patch-size-\$i $>$  pvc-3d-ps\$i.txt\;
done\;

python create-exahype-sycl-plot.py -f pvc-3d-ps4.txt pvc-3d-ps6.txt pvc-3d-ps8.txt -d 3 -ds AoS -ps 4 6 8 -dn PVC\;
  \vspace{0.2cm}
  \caption{
    Run benchmarking executable, pipe the outcome into a text file, and produce
    a graph through matplotlib. \replaced[id=R01]{This example will produce an AoS plot for each implementation plus a comparison of the total runtime and the compute kernel time. The \texttt{ONEAPI\_DEVICE\_SELECTOR} environment variable chooses the target offload device which, in this case, is the first PVC GPU shown after running \texttt{sycl-ls}.}{This example will produce a 2D batched AoS graph.} The text files can be reused for other plots by rerunning the plotting script with different specifications.
    \label{code-snippet:run}
    }
\end{bash}

\replaced[id=R01]{
The directory contains an example SLURM script, \texttt{compile-and-plot.sh}.
}{
The directory hosts two example SLURM scripts
(\texttt{collect-data-and-plot2d.sh} and \texttt{collect-data-and-plot3d.sh}) to run the tests and to produce the output plots.
}
\replaced[id=R01]{
It compiles and
}{
The scripts
}
run the executable and pipe the output into a \texttt{txt} file,
then 
\replaced[id=R01]{
finally feeds
}{they feed} 
the output into a matplotlib script
(\texttt{create-exahype-sycl-plot.py}) to produce the outputs. 
\deleted[id=us]{
Again, the latter accepts numerous alterantive arguments to tweak the visual representation.}
If you prefer to run the examples directly, follow the
instructions~\ref{code-snippet:run}.
\deleted[id=us]{
Our benchmark executable accepts command line arguments---a help message is
displayed if you pass \texttt{--help}--- which
allows users to select an individul SYCL device or to specify how many cores on
the host shall try to offload to the GPU simultaneously (cmp.~work published in \cite{Wille:2023:GPUOffloading}).
}

\section{AoS, SoA and AoSoA}
\label{appendix:further-data}

\deleted[id=us]{Several statements in our results Section~\ref{section:results}
are not supported by data. 
We do so in cases where the additional data would not contribute anything
substantial new to the discussion.
However, some of these additional data are collocated here.
}

% \begin{figure}[htb]
%  \begin{center}
%   \includegraphics[width=0.4\textwidth]{results/A100-NCC-3d/patch-wise-3d-AoS-A100.pdf}
%   \includegraphics[width=0.4\textwidth]{results/pvc-dev-cloud-3d/patch-wise-3d-AoS-PVC.pdf}
%  \end{center}
%  \vspace{-0.6cm}
%  \caption{
%   Cost per degree of freedom update for various $p$ and $T$
%   choices for $d=3$ on an NVIDIA A100 (top) and PVC (bottom). 
%   We employ the patch-wise kernels.
%   \label{figure:results:3d:AoS:patch-wise}
%   \vspace{-0.4cm}
%  }
% \end{figure}

\begin{figure}[htb]
 \begin{center}
  \includegraphics[width=0.4\textwidth]{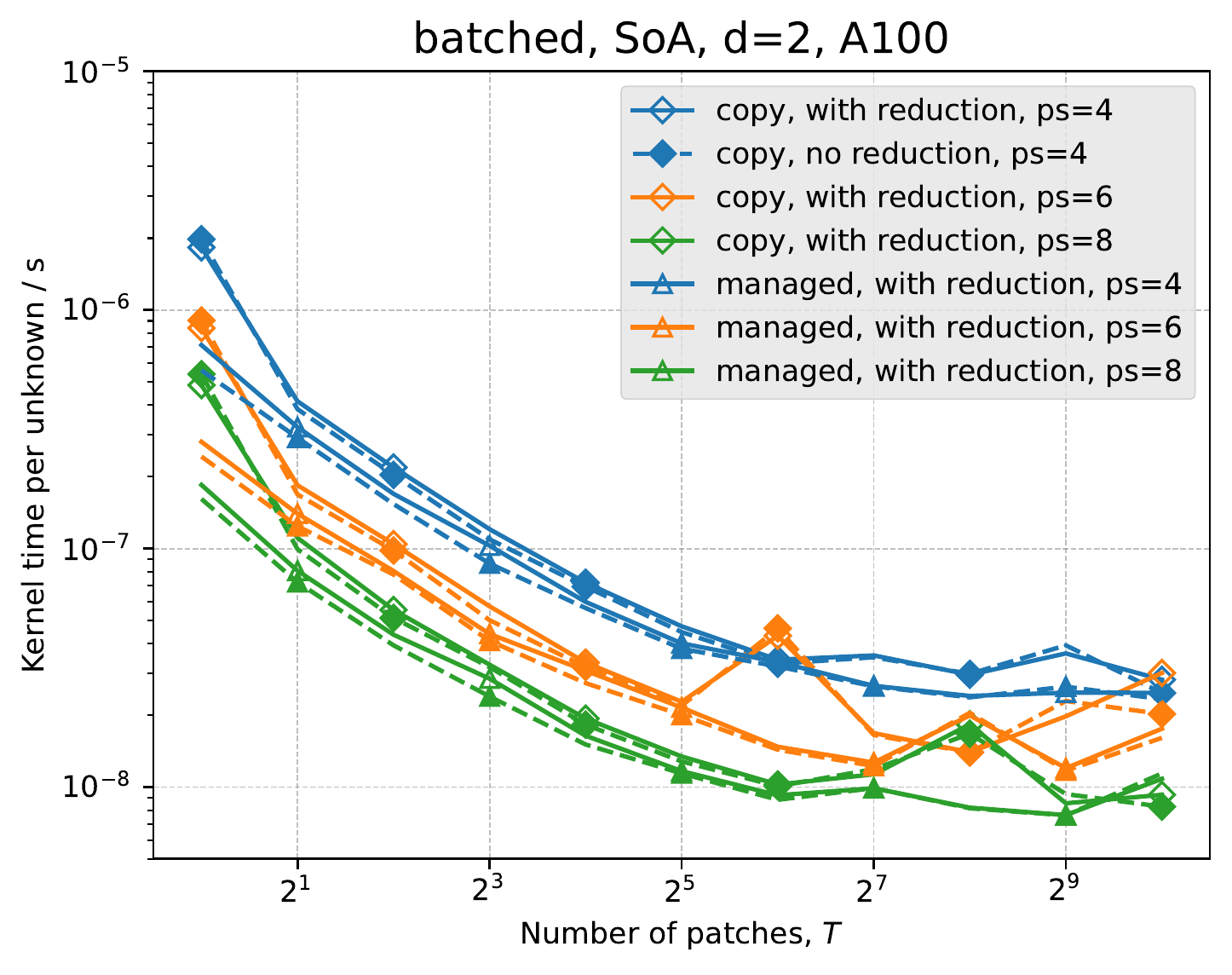}
  \includegraphics[width=0.4\textwidth]{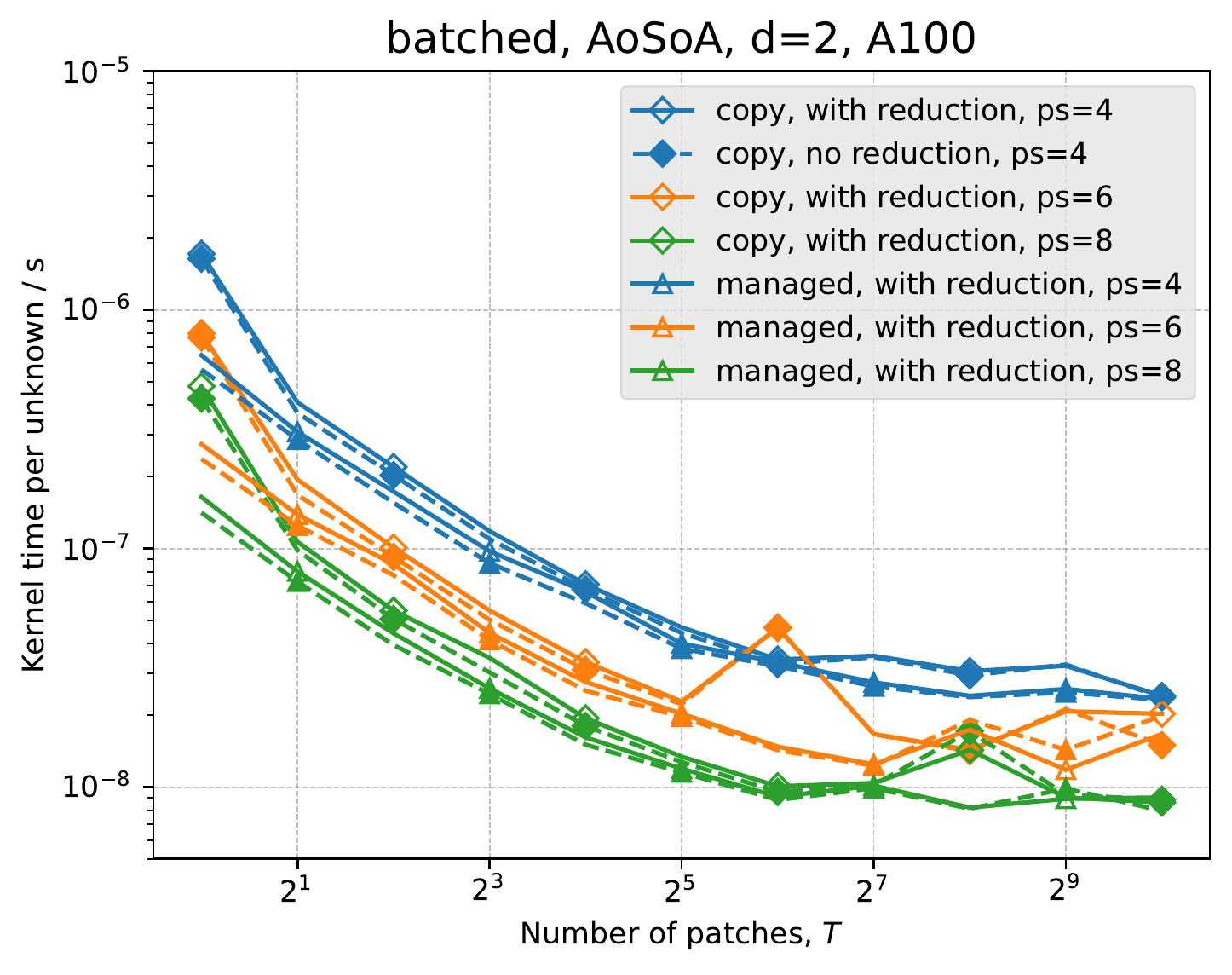}
 \end{center}
 \vspace{-0.6cm}
 \caption{
  $d=2$ results on an NVIDIA A100 using different data layouts within the
  batched kernels.
  \label{figure:results:2d:SoA:A100:batched}
  \vspace{-0.4cm}
 }
\end{figure}

\begin{figure}[htb]
 \begin{center}
  \includegraphics[width=0.4\textwidth]{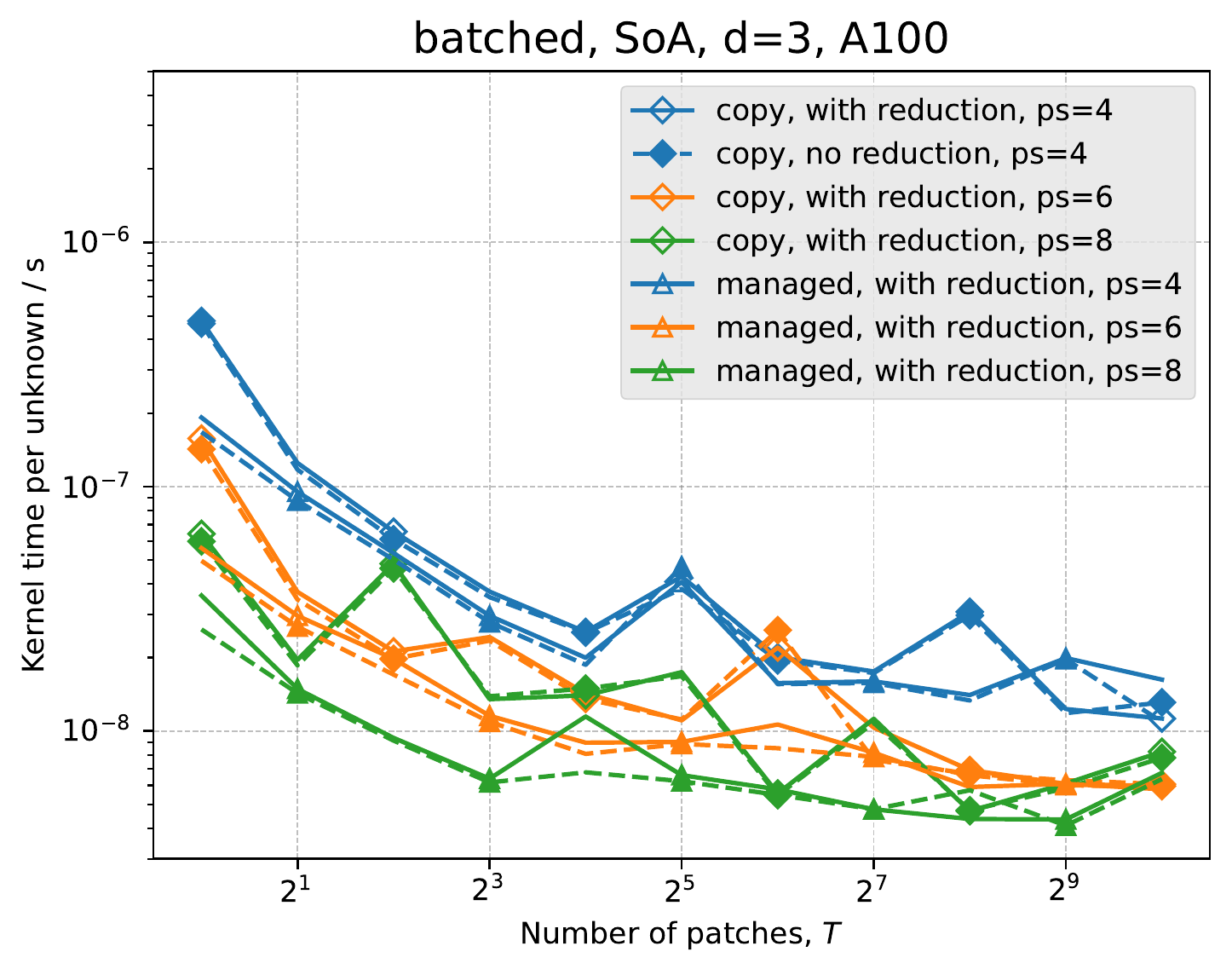}
  \includegraphics[width=0.4\textwidth]{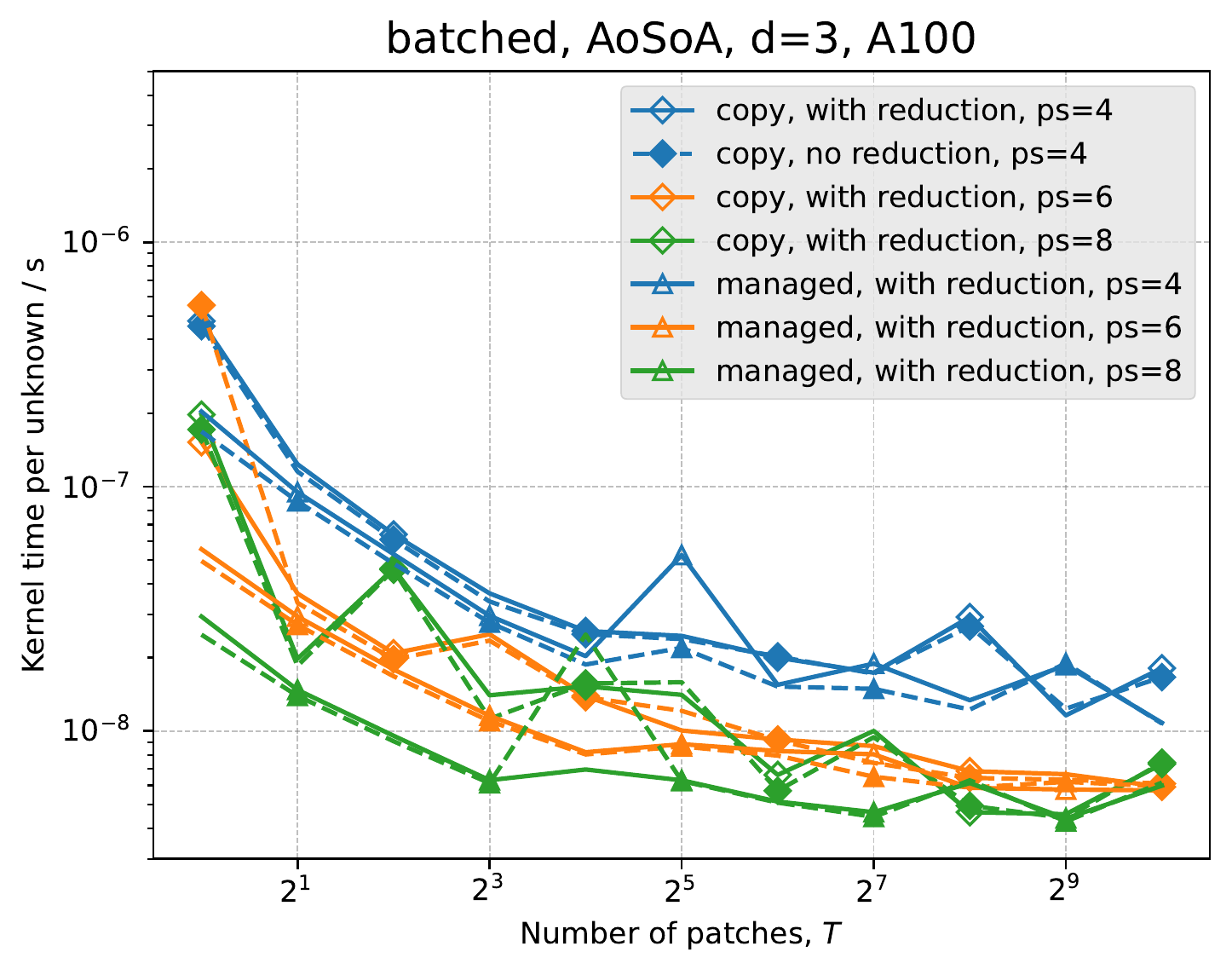}
 \end{center}
 \vspace{-0.6cm}
 \caption{
  $d=3$ results on an NVIDIA A100 using different data layouts within the
  batched kernels.
  \label{figure:results:3d:SoA:A100:batched}
  \vspace{-0.4cm}
 }
\end{figure}

% \begin{figure}[htb]
%  \begin{center}
%   \includegraphics[width=0.4\textwidth]{results/A100-NCC-3d/task-graph-3d-AoS-A100.pdf}
%   \includegraphics[width=0.4\textwidth]{results/pvc-dev-cloud-3d/task-graph-3d-AoS-PVC.pdf}
%  \end{center}
%  \vspace{-0.6cm}
%  \caption{
%  \added[id=us]{@todo}
%   Kernel runtimes on the A100 \added[id=us]{(top) and the PVC (bottom)} for a
%   realisation using a SYCL task graph for $d=2$.
%   \label{figure:results:PVD:task-graph}
%  }
% \end{figure}

\begin{figure}[htb]
 \begin{center}
  \includegraphics[width=0.4\textwidth]{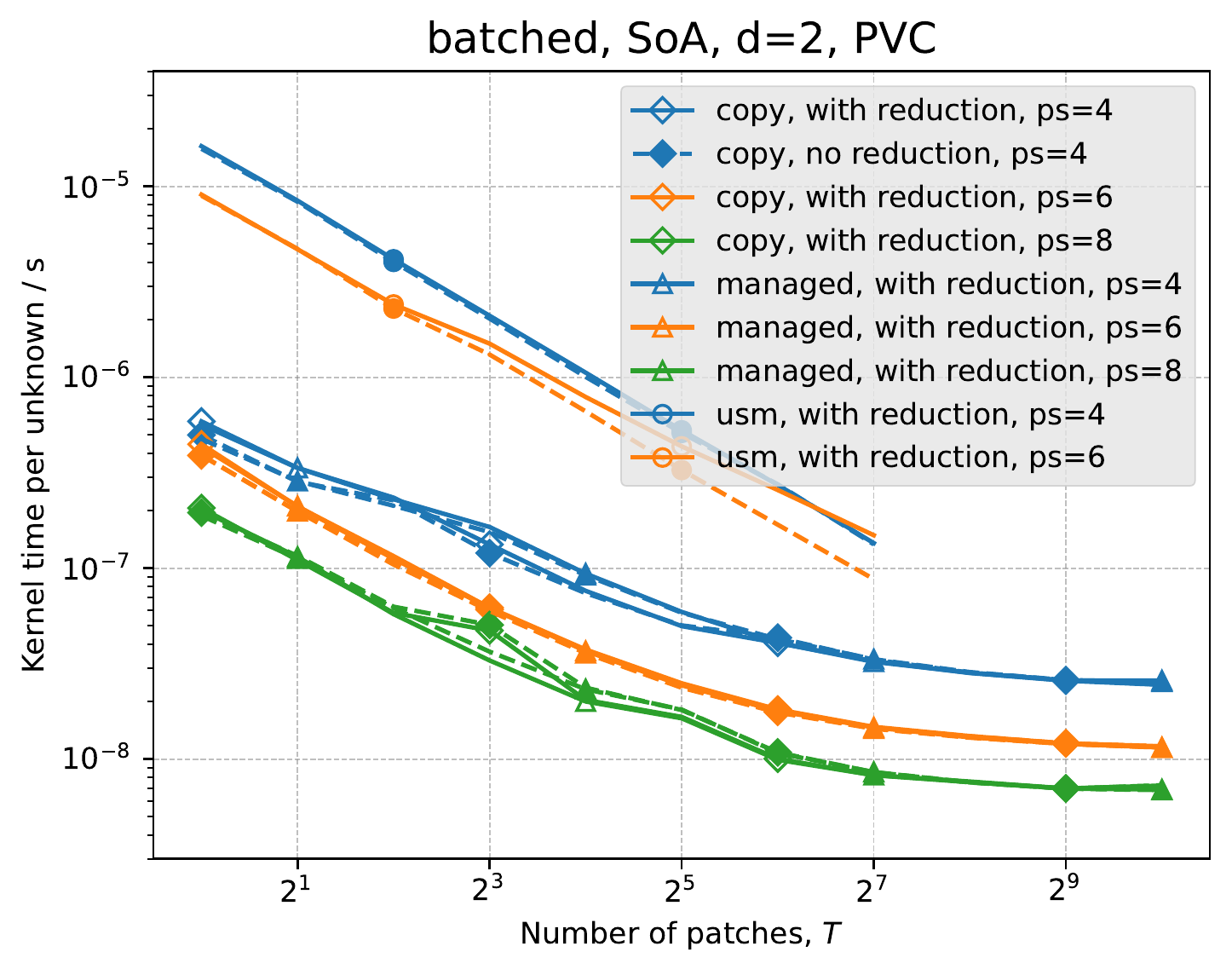}
  \includegraphics[width=0.4\textwidth]{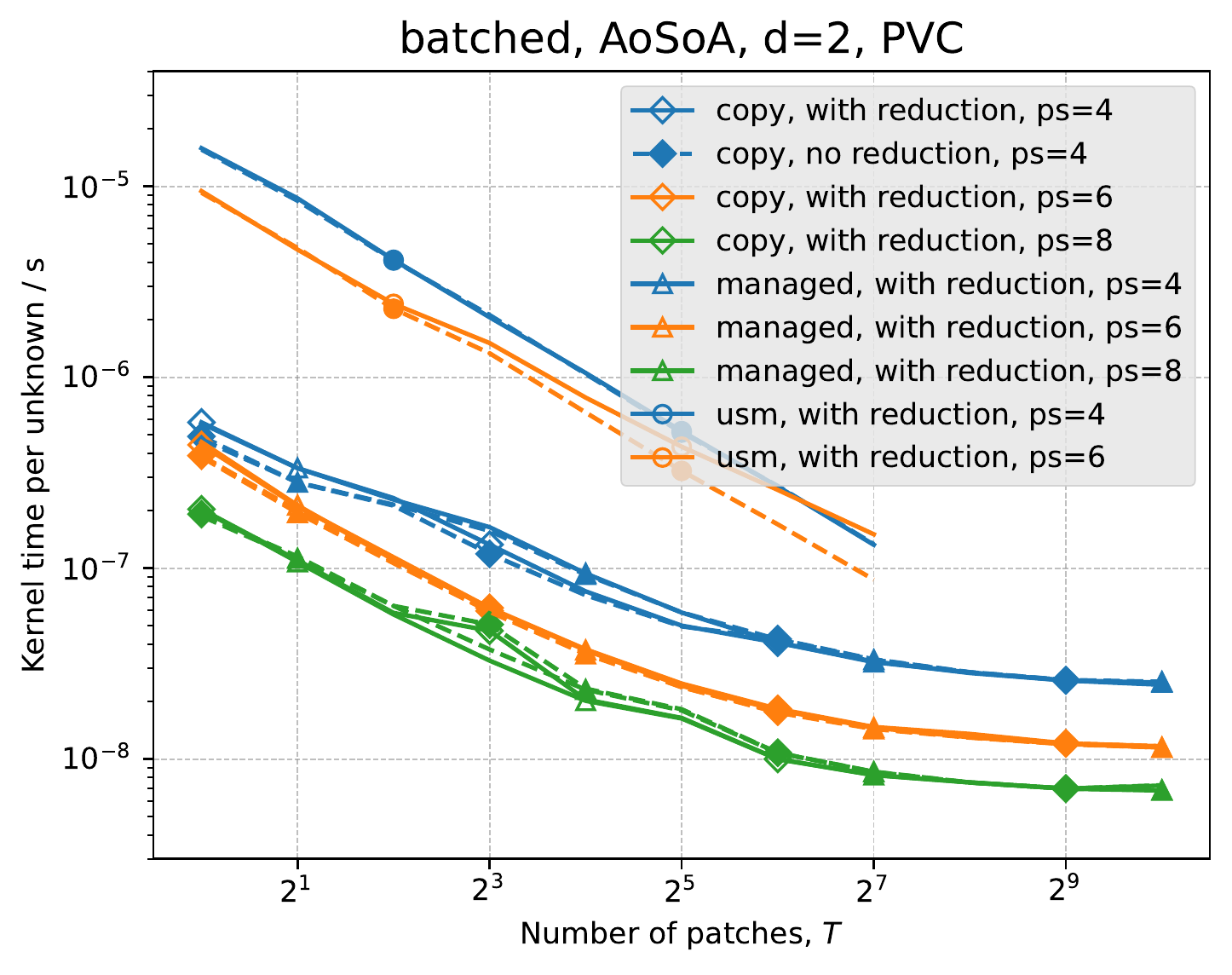}
 \end{center}
 \vspace{-0.6cm}
 \caption{
  $d=2$ results on the PVC using different data layouts within the
  batched kernels.
  \label{figure:results:2d:SoA:PVC:batched}
  \vspace{-0.4cm}
 }
\end{figure}

\begin{figure}[htb]
 \begin{center}
  \includegraphics[width=0.4\textwidth]{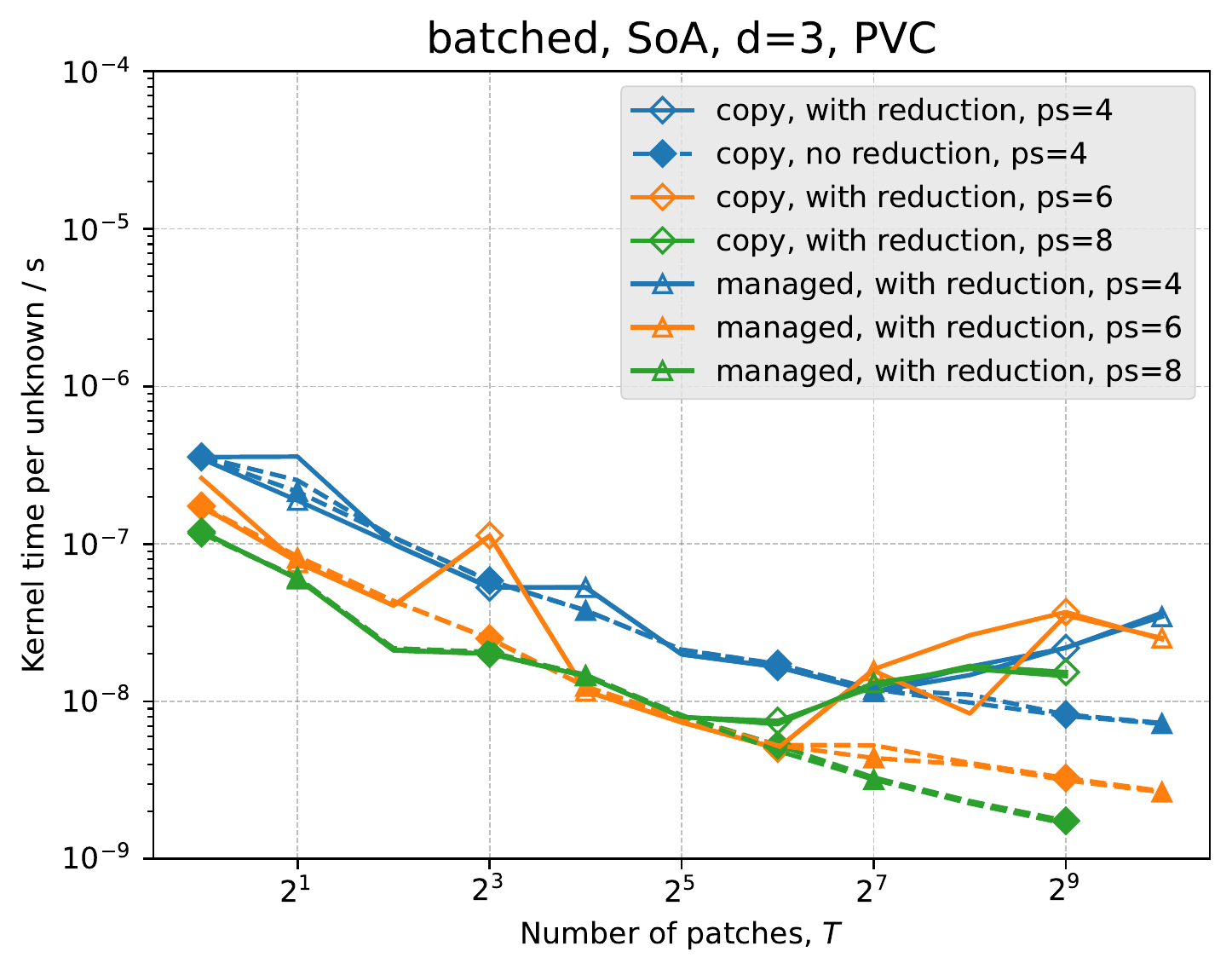}
  \includegraphics[width=0.4\textwidth]{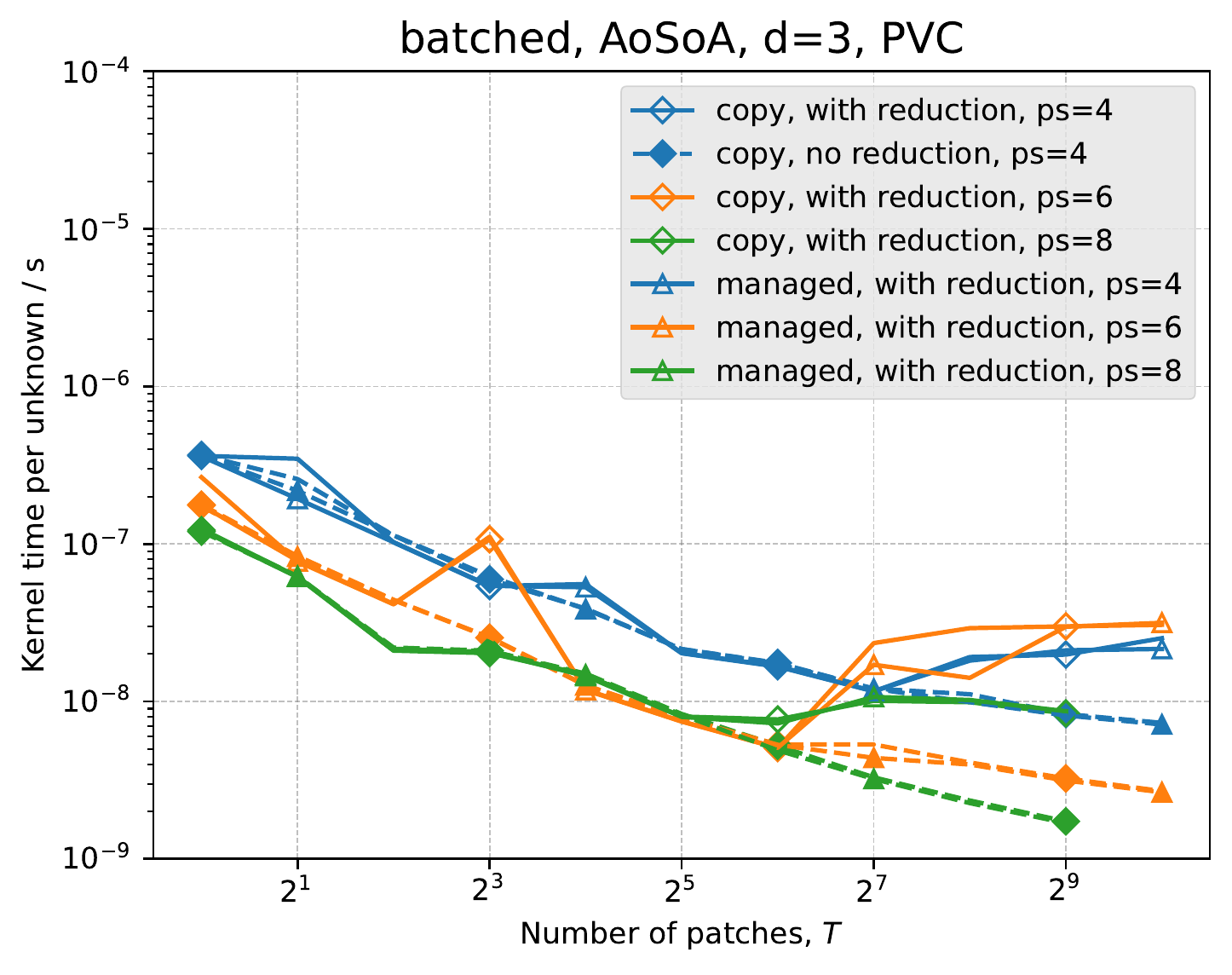}
 \end{center}
 \vspace{-0.6cm}
 \caption{
  $d=3$ results on the PVC using different data layouts within the
  batched kernels.
  \label{figure:results:3d:SoA:PVC:batched}
  \vspace{-0.4cm}
 }
\end{figure}

\replaced[id=us]{Data }{Some data in}  on the runtime impact of AoS vs.~SoA and
a further hybrid (AoSoA) confirms our statement that the organisation of temporary data has negligible
impact on the runtime (Figures~\ref{figure:results:2d:SoA:A100:batched},
\ref{figure:results:3d:SoA:A100:batched},
\ref{figure:results:2d:SoA:PVC:batched},
and \ref{figure:results:3d:SoA:PVC:batched}).
We observe that the characteristic dips for $p=4, d=2$ on the PVC appears
independently of the data layout chosen for the temporary fields.

\section{Machine settings}
\label{appendix:PVC}

\added[id=us]{
 On Intel's Data Center GPU Max Series---abbreviated by PVC in the
 plots, as the previous codename has been Ponte Vecchio---we use the level-Zero API.
 We use only a subset of the available features of the chip
 (Alg.~\ref{algorithm:pvc-settings}).
}

\begin{algorithm}[htb]
 {\footnotesize
export EnableImplicitScaling=0 \;
export ZE\_AFFINITY\_MASK=0.0 \;
export ZE\_FLAT\_DEVICE\_HIERARCHY=FLAT \;
export SYCL\_PI\_LEVEL\_ZERO\_USE\_COPY\_ENGINE=0 \;
 }
 \vspace{0.2cm}
 \caption{
    Environment variables used on PVC.
    \label{algorithm:pvc-settings}
  \vspace{-0.4cm}
 }
\end{algorithm}

\added[id=us]{
 The chip features 1,024 Xe Vector engines organised into two stacks.
 We use only one stack and disable implicit scaling. 
 That is, we do not even allow the chip to make use of the second stack's memory
 and other resources.
 The affinity settings ensure that only this first half of the card is visible
 to the tests, while it is exposed as one flat, uniform device.
}

\added[id=us]{
 Our testbed struggled to benefit from the PVC's copy engine. 
 We therefore explicitly disable this feature throughout the tests, 
 which might negatively affect the USM throughput.
}

\added[id=us]{
 Future driver and software stacks will likely allow users to avoid tinkering
 with environment variables.
 Future hardware and software generations also will enable SYCL USM on the A100. 
 However, we do not expect large quantitative differences in the outcomes and
 notably doubt that the significant performance gap between USM and manual data
 transfer variants can be closed.
}

\added[id=us]{
 For researchers who want to reproduce results, it is important to note that our
 test driver covers all functional variations of the compute kernels.
 If features make the code crash or deadlock, the reproducer benchmark will
 hence not pass either.
 In this case, the respective kernel invocations in the file
 \texttt{KernelBenchmarksFVRusanov-main.cpp} have to be commented out.
}

% Chip besteht aus verschiedenen Tiles, die stacked sind. Die Tiles haben
% verschiedene Funktionen. Uns interessieren nur die Compute Tiles. Den diese
% hosten die Xe Cores.
% 
% Ein Xe Core hat 8 Xe Vector engines.
% Jeder Core hat seinen eigenen L1 Cache.
% 
% 
% Die kleinste Einheit aus der Fabrik ist ein Slice.
% Das ist eine rein organisatorische Einheit, d.h. die Cores sitzen da einfach
% drauf mit ihrem L1 Cache und keinem weiteren Speicher oder so.
% Ein HPC slice has 16 Xe cores. 
% Damit hostet der Slice 16x8=128 Xe Vector engines.
% 
% Die Slices werden dann einen Stack integriert.
% Xe Stack hat 4 Slices und einen eigenen L2 Cache. 
% Also 4x16=64 Xe cores.
% Oder 4x16x8=512 Xe Vector engines.
% 
% Ein Stack hat einen Memory Controller, und einen XeLink.
% Der Xe Link ist im Wesentlichen Aequivalent zu NVLink.
% Stack hat wohl auch eine sogenannte Copy-Engine.
% 
% Xe Link kann bis zu 8 Stacks kombinieren.
% Aber normalerweise nimmt man nur 6 Stacks (6 PVCs) und verwendet die
% verbleibenden zwei XeLink End-Points, um die zum CPU-Speicher zu connecten.
% 
% Den PVC gibt es Liquid (600W) und Thermal (300W) cooled.
% Der 600W hat zwei Stacks.  Also 1024 Xe Vector engines.
% Der 400W hat einen Stack. Aber irgendwie halt nur 56 cores, also 448 Xe engines.	
% 
% Im Level 0 Jargon entspricht die physikalische GPU dem Root Device.
% Wenn es mehr als ein Subdevice gibt, dann sieht Level 0 die so. Allerdings gibt
% es Subdevices eben nur, wenn es mehr als eines gibt. Sonst duerfen sie nicht
% existieren.

\end{document}